# He II Emission from Wolf-Rayet Stars

# as a Tool for Measuring Dust Reddening

Claus Leitherer

*Space Telescope Science Institute[1], 3700 San Martin Drive, Baltimore, MD 21218, USA*

leitherer@stsci.edu

Janice C. Lee

*Caltech-IPAC, MC 314-6, 1200 E California Blvd, Pasadena, CA 91125, USA*

janice@ipac.caltech.edu

Andreas Faisst

*Caltech-IPAC, MC 314-6, 1200 E California Blvd, Pasadena, CA 91125, USA*

afaisst@ipac.caltech.edu

[1] Operated by the Association of Universities for Research in Astronomy, Inc., under NASA contract NAS5−26555

## Abstract

We calibrated a technique to measure dust attenuation in star-forming galaxies. The technique utilizes the stellar-wind lines in Wolf-Rayet stars, which are widely observed in galaxy spectra. The He II 1640 and 4686 features are recombination lines whose ratio is largely determined by atomic physics. Therefore they can serve as a *stellar* dust probe in the same way as the Balmer lines are used as a *nebular* probe. We measured the strength of the He II 1640 line in 97 Wolf-Rayet stars in the Galaxy and the Large Magellanic Cloud. The reddening corrected fluxes follow a tight correlation with a fixed ratio of 7.76 for the He II 1640 to 4686 line ratio. Dust attenuation decreases this ratio. We provide a relation between the stellar $E(B-V)$ and the observed line ratio for several attenuation laws. Combining this technique with the use of the nebular Balmer decrement allows the determination of the stellar and nebular dust attenuation in galaxies and can probe its effects at different stellar age and mass regimes, independently of the initial mass function and the star-formation history. We derived the dust reddening from the He II line fluxes and compared it to the reddening from the Balmer decrement and from the slope of the ultraviolet continuum in two star-forming galaxies. The three methods result in dust attenuations which agree to within the errors. Future application of this technique permits studies of the stellar dust attenuation compared to the nebular attenuation in a representative galaxy sample.

## 1. Introduction

Dust is ubiquitous in star-forming galaxies. Quantifying the amount of dust in galaxies is a prerequisite for determining galaxy properties such as masses, ages, metallicities, star-formation rates and others (Calzetti 2009). The dust attenuation is of paramount importance for understanding spectral energy distributions in the ultraviolet (UV) to near-infrared (IR). The two most widely used techniques to infer the dust reddening compare observations and models of either nebular recombination-line ratios or of the stellar continuum (Calzetti 2013). In this work we discuss an alternative method to determine the dust attenuation in galaxies with active star formation. This method was originally proposed for application to individual stars by Conti & Morris (1990) and makes use of the well-defined properties of Wolf-Rayet (W-R) emission lines observed in the optical and UV spectra of star-forming galaxies.

Conti & Morris (1990) found a tight empirical correlation between the ratio of the He II 1640 ($n = 3 \rightarrow 2$) over He II 4686 ($n = 4 \rightarrow 3$) lines observed in individual W-R stars of spectral type WN. WN stars are the evolved descendants of massive O stars and have ages of a few Myr (Crowther 2007; Massey 2013). Owing to their strong stellar winds, their spectra display mostly broad ($\sim 10^3$ km s$^{-1}$) emission lines, with the He II lines being sign-posts commonly observed in the integrated spectra of local star-forming galaxies (Kunth & Sargent 1981; Conti 1991; Schaerer et al 1999; Brinchmann et al. 2008; López-Sánchez & Esteban 2008; Miralles-Caballero et al. 2016) and of Lyman-break galaxies at redshifts up to and beyond $z \approx 3$ (Shapley et al. 2003; Erb et al. 2010; Shapley 2011; Faisst et al. 2016; Steidel et al. 2016; Rigby et al. 2018). The work of Conti

& Morris established a *stellar* technique analogous to the *nebular* recombination-line method: two spectral lines widely separated in wavelength have a fixed flux ratio whose deviation from the intrinsic value indicates the reddening. The physical reason is understood: He II is close to a pure recombination line in W-R winds. In the optically thin case, the predicted ratio of the He II 1640 over 4686 lines would be around 8, increasing both with electron temperature and density (Hummer & Storey 1987; see Section 4). Radiative transfer effects and measurement uncertainties due to the contribution from other blended lines (like C III/IV) may affect the Case B value. Conti & Morris performed a first test and determined the reddening of a sample of 30 Galactic W-R stars using the He II line ratio. The comparison with the reddening obtained independently from a comparison of the observed and theoretical continuum supported the promise of this method. The limitation of that study was the availability of UV spectra. At that time only a limited number of UV spectra of W-R stars collected with the International Ultraviolet Explorer (IUE) satellite was available. Equally important, IUE was the only UV spectroscopic mission at that time but did not have the sensitivity for high-S/N spectroscopy of galaxies. As a result, the technique of using the He II lines in W-R stars has never been fully calibrated and applied to the integrated spectra of galaxies displaying W-R signatures.

Motivated by the approach of Conti & Morris (1990), we identified all suitable UV spectra of W-R stars collected by various space missions and increased their original sample by almost a factor of two (Section 2). The data analysis is described in Section 3, where we also compare our sample of W-R stars to that of Conti & Morris. In this section we also discuss measurements of the optical He II 4686 line, which were retrieved from the literature. The empirical relation between the equivalent widths of the He II 1640 and 4686 is presented in Section 4. Section 5 describes the correlation between the lines fluxes of the two lines and gives practical recipes from deriving the reddening in galaxies for several reddening laws. Section 6 presents a first application of the method to star clusters in nearby galaxies and compares the results to the reddening derived from the UV continuum slope and from the Balmer decrement. The potential of this technique and future applications are discussed in Section 7.

## 2. Program Star Selection

Conti & Morris (1990) included 53 Galactic and LMC W-R stars in their analysis, which represented the complete sample of UV spectra available to them at that time. Since their study had been completed, the number of W-R spectra collected with various space missions has increased considerably. We searched the Mikulski Archive for Space Telescopes (MAST) for all UV spectra of W-R stars of the nitrogen sequence in the Galaxy, the LMC, the Small Magellanic Cloud (SMC), M31, and M33. W-R stars of type WC were not considered in our query. The relevant missions are the Hubble Space Telescope (HST), IUE, and the Hopkins Ultraviolet Telescope (HUT). W-R spectra obtained by the Far-Ultraviolet Spectroscopic Explorer (FUSE) were located as well, but the spectral range of FUSE does not extend to He II 1640. We identified six suitable spectra of WN stars in the SMC, three of which are known binaries (AB 3, AB 6, AB 7;

Shenar et al. 2016). The remaining three stars (AB 1, AB 2, AB 4) had useful data but were subsequently discarded because of the trade-off between increasing the metallicity range versus adding three stars to an otherwise homogeneous sample in terms of chemical composition. The spectra of all W-R stars in M31 and M33 had insufficient signal-to-noise for our study.

Spectra of Galactic and LMC W-R stars were collected by HST, IUE and HUT. HUT observed five Galactic W-R stars (Schulte-Ladbeck et al. 1995) and three W-R-related stars in the LMC (Walborn et al. 1995). All eight stars were observed by IUE as well, and the data quality of the IUE spectra is comparable to, or better than that of HUT. Therefore we did not consider the HUT spectra any further. As for HST data, MAST contains three Galactic W-R spectra covering He II 1640, and spectra of 19 unique LMC stars. 14 of the 19 LMC stars were observed with HST's first-generation spectrographs Goddard High-Resolution Spectrograph (GHRS) and Faint Object Spectrograph (FOS). The remaining Galactic and LMC stars have data obtained with the Space Telescope Imaging Spectrograph (STIS). All Galactic and LMC stars were observed with IUE as well, and the quality of the IUE data is comparable to that of the HST data. Since the IUE data include all candidate stars, we decided to exclusively utilize IUE data for the sake of homogeneity. Our total sample contains 45 Galactic and 52 LMC W-R stars with IUE spectra. Table 1 summarizes the Galactic program stars. Column 1 of this table gives the W-R ID used by van der Hucht (2001). An alternate designation is in column 2. Right ascension and declination (van der Hucht) are in columns 3 and 4, respectively. These coordinates are not necessarily identical to those attached to the star in MAST, as the observer may have entered different coordinates. The MAST coordinates are pre-observation and are not updated to the final position after target acquisition. Column 5 of Table 1 gives the spectral types from van der Hucht. The classification largely follows the scheme of Smith (1968a), with the amendments and references given by van der Hucht. The corresponding information for the LMC W-R stars is provided in Table 2. Here the W-R ID and alternate ID in columns 1 and 2, respectively, are from Breysacher et al. (1999). The right ascension, declination, and spectral types in columns 3, 4, and 5, respectively, are from the same source. Individual references for the spectral types can be found in Breysacher et al. Our sample of 97 W-R stars nearly doubles the sample originally available to Conti & Morris (1990). Almost all program stars analyzed by Conti & Morris are included in our final data set.

## 3. Analysis of the UV Spectra and Ancillary Optical Data

We retrieved all available spectra of the 97 W-R stars listed in Table 1 and Table 2 from MAST. The spectra were obtained with IUE's SWP camera in the low-dispersion mode at a resolution of approximately 6 Å. The 10'' × 20'' entrance aperture was used for all considered spectra in order to maximize the photometric accuracy. Instrumental details of the IUE satellite and its spectrograph can be found in Boggess et al. (1978a, b). Spectra whose region around He II 1640 was flagged as having saturated pixels were eliminated from the sample and not considered. In cases when more than one spectrum was available for the same program star, we used the spectrum with the highest S/N for the measurements. Column 6 of Table 1 and Table 2 identifies the SWP spectrum used for the Galactic and LMC stars, respectively. We measured the equivalent widths *EW*(1640) and line fluxes *F*(1640) by integrating the observed profiles of He II 1640 over a wavelength range of 30 Å. The central wavelength was 1640.4 Å for the Galactic stars, and 1641.8 Å for the LMC stars (corresponding to an LMC recession velocity of 262 km $s^{-1}$). The measurements are in columns 7 and 8 of Table 1 and Table 2. The measured values of *EW*(1640) range between about 5 and 150 Å. The principal uncertainty is the placement of the adopted continuum level. We determined the continua by inspecting the full wavelength range, manually choosing regions between emission lines as representative of the stellar continuum, and interpolating between these regions to fit the full continuum. Conti & Morris (1990) estimate errors of 30% for their measurements. Since their data and ours are similar, we assume the same errors for our measurements as well. As a test of the accuracy of our measurements, we performed a comparison between our results and those of Conti & Morris, who have 50 stars in common with our sample. In Figure 1 we show the comparison of *EW*(1640) between the measurements obtained here and those of Conti & Morris for the 50 Galactic and LMC stars. The error bar in this figure corresponds to a 30% uncertainty in both our and the Conti & Morris data set. We find excellent agreement with a mean difference of ($-0.04 \pm 0.02$) Å between log(*EW*(1640)) measured by us and by those authors. In order to identify offsets between the Galactic and the LMC sample we show in Figure 2 the ratio of the two measurements as a function of the measurements in this work. The mean differences for the Galactic and the LMC stars are the same: ($-0.04 \pm$ 0.02) Å for either stellar group. The corresponding tests for *F*(1640) are in Figure 3 and Figure 4. In this case the logarithmic difference is (0.01 ± 0.02) erg $s^{-1}$ $cm^{-2}$ for the full sample. The mean difference for the Galactic stars is (0.02 ± 0.02) erg $s^{-1}$ $cm^{-2}$, and for the LMC stars it is (-0.01 ± 0.02) erg $s^{-1}$ $cm^{-2}$. The comparison suggests no systematic offset between the results of Conti & Morris and those derived here.

We searched the literature for published measurements of the optical He II 4686 line in our 97 program stars with available He II 1640 data. The results of the literature search are in Table 3. This table gives the W-R ID (column 1), the published equivalent width of the He II 4686 line *EW*(4686) (column 2), the line fluxes *F*(4686) (column 3), and the references for the data (column4). In cases of multiple measurements, we took the average of all data. While *EW*(4686) is available for all program stars, some stars lack published *F*(4686) values as indicated by the lack of entries in column 3 of Table 3. We recovered the missing fluxes by utilizing existing continuum photometry. For this purpose we compiled *v* (column 5) and $(b-v)$ (column 6) published in the literature (column 7). The *ubv* narrow-band photometric system was devised by Smith (1968b) to target the line-free continuum of W-R stars. The resulting magnitudes correspond to *UBV* magnitudes corrected for the presence of emission lines. For the few cases with no $(b-v)$ literature values but available $(B-V)$ broad-band colors, we used the relation

$$(b-v) = 0.78\,(B-V) - 0.09 \qquad (1)$$

by Lundström & Stenholm (1980) to calculate $(b-v)$. We then transformed the *b* and *v* magnitudes into absolute fluxes following Schmutz & Vacca (1991), whose calibration leads to absolute fluxes at zero magnitude of $f(b) = 5.55 \times 10^{-9}$ erg $s^{-1}$ $cm^{-2}$ $Å^{-1}$ and $f(v)$ =

$3.91 \times 10^{-9}$ erg s$^{-1}$ cm$^{-2}$ Å$^{-1}$. Using this calibration we obtained continuum fluxes at *b* and *v*, which were then interpolated to a wavelength of 4686 Å and then multiplied by *EW*(4686) in column 2 of Table 3. The resulting calculated He II 4686 line fluxes $F_{calc}$(4686) are in column 8 of the table.

We tested the reliability of this method by comparing $F_{calc}$(4686) to *F*(4686) of those stars for which the line fluxes had been measured. The outcome of this test is shown in Figure 5, where we have plotted the observed versus the calculated values along the abscissa and ordinate, respectively. The 74 stars in this figure follow a tight relation with an average log($F_{calc}$ (4686)) – log(*F* (4686)) = (–0.02 $\pm$ 0.02) erg s$^{-1}$ cm$^{-2}$. We broke the full sample into two subsamples consisting of the Galactic and LMC stars and plotted the logarithmic ratios of log($F_{calc}$(4686)) – log(*F*(4686)) in Figure 6. The mean differences for the Galactic and LMC W-R stars are (0.03 $\pm$ 0.02) erg s$^{-1}$ cm$^{-2}$ and (–0.06 $\pm$ 0.02) erg s$^{-1}$ cm$^{-2}$, respectively. The values of the two subsamples agree within the errors. The excellent agreement between *F*(4686) and $F_{calc}$(4686) gives confidence in the calculated values, which are used to fill in the missing entries in column 3 of Table 3. Consequently our sample has 97 stars with observed *EW*(4686), 74 of which have observed *F*(4686), and 23 have $F_{calc}$(4686). In the following we will substitute $F_{calc}$(4686) for the missing measurements of the 23 W-R stars and refer to their fluxes values as *F*(4686).

## 4. Relation between He II 1640 and 4686

In this section we compare the equivalent widths and line fluxes of the UV and optical He II lines. Figure 7 shows the relation between *EW*(1640) (column 7 of Table 1 and Table 2) and *EW*(4686) (column 2 of Table 3). Here we assume that the data for the optical He II 4686 line have the same error as the UV data, i.e., 30%. The figure suggests a significant correlation between the two He II lines, with several striking outliers. By definition, the equivalent widths of spectral lines in individual stars are reddening- and distance independent. Any scatter around the relation is due to measurement errors, stellar variability (as the two measurements are not contemporaneous), variations of the line ratio with W-R spectral type, and – most importantly – the presence of a secondary star affecting the continuum. The most deviant data points are at $\log(EW(4686)) \lesssim 0.5$. Inspection of Table 2 indicates that all four stars have spectral types later than WN8, i.e., WN9, WN10, and WN11. We identified all program stars with these spectral types and flagged them in Figure 7. There are eight stars: WR 108 (WN9), BAT99 13 (WN10), BAT99 22 (WN9), BAT99 33 (WN9/Ofpe), BAT99 55 (WN11), BAT99 76 (WN9), BAT99 120 (WN9), and BAT99 133 (WN11). W-R stars of the latest spectral types resemble extremely luminous O stars in their spectral morphology and are most likely hydrogen-rich stars still on the main-sequence (Crowther & Bohannan 1997). They are cooler than their counterparts with earlier spectral types, and any O-star companion will have a stronger diluting effect on the equivalent widths. This is expected to affect the optical He II line more than the UV line because the companion is less luminous in the UV than the hotter W-R star. Furthermore, He II is no longer a pure emission line but develops a P Cygni profile in WN9 – 11 stars, and radiative transfer effects become important. Therefore we eliminated the eight named stars from further analysis. A least-square fit to the remaining 89 Galactic and LMC stars gives

$$EW(4686) = 0.97\ EW(1640) + 0.58 \tag{2}$$

suggesting that the ratio between the two equivalent widths is almost independent of the stellar properties and has a value close to 4. Log(*EW*(4686)) – log(*EW*(1640)) is shown in Figure 8. The mean logarithmic differences for the Galactic and LMC sample are (0.63 ± 0.03) Å and (0.43 ± 0.03) Å, respectively. The value for the combined sample is (0.54 ± 0.03) Å. The slight offset may indicate a real difference in the equivalent widths of the two samples. Since the samples were constructed from rather inhomogeneous sources, there is no obvious explanation. A similar offset is seen in the sample analyzed by Conti & Morris (1990; their Figure 2).

Next we consider the line fluxes. They are plotted in Figure 9, which has the same structure as the previous figure. The eight W-R stars with anomalous equivalent-width ratios are again included in the plot and flagged. The line fluxes are not affected by the presence of a companion unless the companion has strong He II lines itself. Several effects shape the distribution of the data points in Figure 9. Any dependence of the ratio of the two He II lines on stellar properties (which we will show later to be insignificant) would weaken the correlation, as will measurement errors. The spread of the data points from the lower left to the upper right is caused by the dependence of the line strength on the W-R properties, the distance of the stars, and the reddening. Strikingly, the Galactic and LMC stars form two distinct population in the figure. The Galactic stars extend over a larger range of line fluxes, and the slope of their flux relation differs from that of the LMC stars. We have included the reddening vector for *E(b−v)* = 0.5 in Figure 9. This vector results from the Galactic law of Mathis (1990), which predicts a total absorption ratio at 1640 and 4686 Å of *A*(1640)/*A*(4686) = 2.09. The distribution of the Galactic stars is parallel to the reddening vector, suggesting their location is largely determined by reddening. In contrast, most LMC stars have much lower reddening, and therefore their location in Figure 9 reflects the intrinsic line ratio.

We corrected *F*(1640) and *F*(4686) for reddening using published values of *E(b−v)*. The adopted values and their references are in columns 2 and 3 of Table 4, respectively. The majority of the color excesses are from the comprehensive surveys of Hamann et al. (2006) for the Milky Way stars and of Hainich et al. (2014) for the LMC stars. In both studies the W-R properties were derived by modeling the line spectrum with theoretical atmospheres, and the dust reddening was determined from a comparison of the theoretical and observed colors. For a few stars not included in Hamann et al. and Hainich et al. we adopted the *E(b−v)* values of Schmutz & Vacca (1991) and Crowther & Dessart (1998). Where applicable, the conversion from *E(B−V)* to *E(b−v)* was done with the relation *E(b−v)* = 0.83 *E(B−V)* (Lundström & Stenholm 1980). The *E(b−v)* values were used to correct *F*(1640) and *F*(4686) for reddening choosing the reddening laws of Mathis (1990) and Fitzpatrick (1986) for the Galactic and LMC W-R stars, respectively. The reddening-free line fluxes $F_0(1640)$ and $F_0(4686)$ are in columns 4 and 5 of Table 4.

We define the intrinsic ratio of the He II 1640 and 4686 lines as

$$R_0 = \frac{F_0(1640)}{F_0(4686)}. \qquad (3)$$

The calculated $R_0$ values for all 89 stars are listed in column 6 of Table 4. Inspection of column 6 as well deviations of the individual values from the calculated mean suggest two significant outliers: WR 25 and WR 43a. WR 25 is a known binary and has recently be reclassified as O2.5If*/WN6 by Crowther & Walborn (2011), suggesting a close relationship to very massive O stars. WR 43a is a double-eclipsing binary system located in the NGC 3603 cluster (Moffat et al. 2004). It has a period of 3.77 d, and exhibits brightness variations of about 0.3 mag, making it likely that the reason for the anomalous $R_0$ are the non-contemporaneous line-flux measurements. Therefore, we removed both stars from our list so that the final sample consists of 87 stars. A plot of the dereddened He II 1640 and 4686 line fluxes for is reproduced in Figure 10. In the figure we flagged the eliminated stars WR 25 and WR 43a and omitted the eight WN9 – 11 stars previously excluded from consideration. Comparison with Figure 9 suggests that reddening is indeed responsible for the offset between the Galactic and LMC stars in the prior figure. After reddening correction, both the Galactic and the LMC stars have smaller scatter around the mean and follow the same relation. The separation of the two populations results from the larger distance of the LMC stars, and the spread within each population reflects the varying line strengths of both lines (*not the ratio*) with spectral type. A least-square fit to the data points in Figure 10 (excluding WR 25 and WR 43a and the WN9 – 11 stars) gives

$$\log F_0(4686) = (0.95 \pm 0.02) \log F_0(1640) + (-1.36 \pm 0.17). \qquad (4)$$

The fit suggests a small (0.95 $\pm$ 0.02) but not significant trend of $R_0$ with the line fluxes. This is most likely the result of slightly different wind properties of Galactic and LMC stars. We performed least-square fits to the Galactic and LMC stars separately. Then the values for the slopes and intercepts in equation (4) become 0.92 ± 0.05 and –1.62 ± 0.44 for the Galactic stars, respectively. The LMC values are 0.75 ± 0.06 and –3.60 ± 0.66, respectively. The uncertainties of the fit coefficient are larger due to fewer data points. In particular the most deviant data points of the LMC sample are at the extreme faint end in Figure 10. Excluding the two most extreme values would change the slope from 0.75 to 0.82 and the intercept from –3.60 to –2.86. We conclude that the dispersion around the regression line and the offset between the samples can be understood solely in terms of the data uncertainties, and we will not divide the sample. In order to highlight any differences of the properties of the two subsamples we plotted the logarithmic differences log $F_0$(4686) – log $F_0$(1640) for both the Galactic and the LMC stars in Figure 11. The differences are (0.93 ± 0.02) erg $s^{-1}$ $cm^{-2}$ and (0.86 ± 0.03) erg $s^{-1}$ $cm^{-2}$ for the Galactic and LMC stars, respectively. The combined value is (0.89 ± 0.03) erg $s^{-1}$ $cm^{-2}$, which is not significantly different from the two individual values. There are two error sources: the flux measurement error and the adopted *E(b–v)*. Taking the previously adopted measurement uncertainties of 30% as indicative of the errors, we expect typical errors of $\pm$ 0.12. This value is plotted as the error bar in Figure 10. The literature sources for *E(b–v)* do not give errors. If we assume typical uncertainties of 0.05 mag, the error introduced by the reddening correction would be around $\pm$ 0.2, with the correction in the UV dominating. In total, we expect the errors of the individual data point in Figure 10 to be about ± 0.25, which implies that 82% of the values agree within their errors with the mean relation. We will therefore adopt log $R_0$ = 0.89 or $R_0$ = 7.76 as the intrinsic value of the ratio of the two helium lines.

How does this value compare to other independent determinations? Schaerer & Vacca (1998) compiled observed line ratios from various literature sources which overlap with our sample. They found average line ratios of 7.55 and 7.95 for late WN stars in the Galaxy and the LMC, respectively. Stellar atmosphere models suggest a ratio of order 10 for all WN types (Crowther & Hadfield 2006). Hummer & Storey (1987) computed a large grid of line ratios of hydrogen-like ions over a wide range of electron densities and temperatures. Taking $n_e$ = $10^9$ cm$^{-3}$ and $T_e$ = 20,000 K as representative for W-R winds, their Case B value = 8.36.

## 5. Reddening from the 1640/4686 Ratio

The result of a fixed ratio of the two He II lines for WN stars permits a yet unexplored method to determine the dust attenuation from an analysis of the integrated spectra of star-forming galaxies with detected W-R features. Dust will decrease the intrinsic value of $R_0$. Comparison between the observed ratio $R_{obs}$ and $R_0$ provides an estimate of the stellar dust attenuation. We can quantify to relation between $R_{obs}$ and the corresponding reddening *E(B−V)* for several commonly used laws. The reddened and the intrinsic line ratios are related to the dust attenuation by

$$\frac{R_{obs}}{R_0} = 10^{0.4\,E(B-V)\,(k(4686)-k(1640))}, \quad (5)$$

where $k$(1640) and $k$(4686) are the absorption coefficients at 1640 and 4686 Å, respectively. Equation (5) can be rewritten to

$$E(B-V) = \frac{1}{0.4\,(k(4686)-k(1640))} \log R_{obs} + \frac{0.89}{0.4\,(k(1640)-k(4686))}. \quad (6)$$

Here we have inserted the derived intrinsic line ratio of log $R_0$ = 0.89. We define the zeroth and first order coefficients as A and B, so that equation (6) becomes

$$E(B-V) = \mathrm{A} \log R_{obs} + \mathrm{B}. \quad (7)$$

A and B depend on the choice of the reddening law. We consider five different reddening laws: the Milky Way law of Mathis (1990), the extinction laws of Gordon et al. (2003) derived for LMC and SMC stars, the extragalactic attenuation law of Calzetti (2001) for local star-forming galaxies, and the attenuation law determined by Reddy et al. (2015) for galaxies at redshift 1.4 to 2.6 in the MOSDEF survey. The Calzetti and Reddy laws result in very similar reddening corrections. These reddening laws are summarized in Table 5, where we give the ratios of the total to selective absorption $R_V$ used in the five laws (column 2), and the values of $k$(1640) (column 3) and $k$(4686) (column 4). The two absorption coefficients were calculated as follows. Mathis and Gordon et al. parameterize their extinction laws in terms of the total absorption at wavelength λ, $A$(λ), relative to the total absorption at a fixed wavelength. We obtained $k$(λ) from the relation

$$k(\lambda) = \frac{A(\lambda)}{A(V)} R_V \quad (8)$$

using the $R_V$ values in Table 5. Calzetti and Reddy et al. approximate their reddening curves as polynomial fits to $k(\lambda)$ and we are adopting their equations directly. Calzetti et al. (2000) provide this relation:

$$k(\lambda) = 2.659\ (-2.156 + \frac{1.509}{\lambda} - \frac{0.198}{\lambda^2} + \frac{0.011}{\lambda^3}) + 4.05. \quad (9)$$

The corresponding equation of Reddy et al. is

$$k(\lambda) = 5.729 + \frac{4.004}{\lambda} - \frac{0.525}{\lambda^2} + \frac{0.029}{\lambda^3} + 2.505. \quad (10)$$

Equations (8), (9), and (10) were used to calculate A and B in columns 5 and 6 of the table, respectively. Equation (7) and a choice of A and B allow one to use the observed ratio of the He II 1640 and 4686 lines in galaxies with W-R features to determine the dust attenuation of the stellar population. There are three sources of errors determining $E(B-V)$ from equation (7). First of all, a choice for the reddening law must be made. The two extreme examples in Table 5 are the Milky Way law, which is very shallow, and the SMC law, which is very steep. For an observed value of $R_{obs}$ = 1, the former predicts $E(B-V)$ = 0.53 and the latter $E(B-V)$ = 0.28. The widely used extragalactic attenuation laws of Calzetti et al. and Reddy et al. result in $E(B-V)$ = 0.44. These two extragalactic relations may be most appropriate for the reddening correction of galaxy SEDs. However, the appropriate attenuation law will depend on galaxy type, chemical composition, redshift, as well as other parameters (Salim et al. 2018). One should also keep in mind that the reddening law within a galaxy itself may be variable (Decleir et al. 2019).

Equation (7) was derived for our measured value of log $R_0$ = 0.89, for which we estimated an uncertainty of $\pm$ 0.1. This then translates into an error of approximately $\pm$ 0.05 for *E(B−V)* for all reddening laws. Finally we need to consider the error in the measured line ratio of the two helium lines. If the flux of each line has an error of 20%, the derived *E(B−V)* from equation (7) has an uncertainty of about $\pm$ 0.1. We can compare this error with that associated with deriving the stellar dust attenuation in galaxies from the UV spectral slope β. Using the relation β = –2.44 + 4.54 *E(B−V)* of Reddy et al. (2015) and their estimated uncertainty of Δβ = 0.2, we find an error of $\pm$ 0.1 for *E(B−V)*. Therefore, determining the dust reddening from the UV continuum or the ratio of the He II 1640 and 4686 lines is expected to have similar error bars for *E(B−V)*.

## 6. Application to Star Clusters in Nearby Galaxies

As a first test of the method for integrated stellar populations, we measured the He II lines from star clusters in the nearby galaxies NGC 3049 (*D* = 11.6 Mpc; McQuinn et al. 2017) and NGC 5398 (*D* = 14.7 Mpc; Sidoli et al. 2006). These galaxies are chosen because they have previously been classified as W-R galaxies from their optical spectra (Schaerer et al. 1999), and bracket a large range in oxygen abundance -- from 0.25 $Z_\odot$ (NGC 5398) to 1.2 $Z_\odot$ (NGC 3049). Most importantly, the galaxies have co-spatial UV-optical spectra obtained with STIS available in the MAST archive, which allows for the calculation of the three dust attenuation indicators He II line ratio, Balmer line ratio, and UV continuum slope.

The STIS data for both galaxies were taken as part of HST program 7513 (PI: C. Leitherer) in 1999 – 2000. HST spectroscopy was obtained using STIS both in the UV (FUV-MAMA; G140L grating; 52" × 0.5" aperture) and optical (CCD; G430L & G750M gratings; 52" × 0.1" aperture). The data for NGC 3049 have been published in González Delgado et al. (2002), and for NGC 5398 in Sidoli et al. (2006), and the reader is referred to those papers for details regarding the observations. Here we perform an independent standard reduction of the spectra to enable the measurements necessary for consistent analysis of the reddening with all three indicators. It should be noted that the optical STIS spectra are shallow because the original program was focused on strong emission lines, and was not optimized for the measurement of broad He II 4686 line. We first briefly discuss key aspects of our measurement procedure and reddening calculation, and then present the suite of results for each galaxy. The spectra are shown in Figure 12, and all measurements and results are summarized in Table 6.

We measured the UV continuum slope β by performing a linear least-squares fit to the data between 1240 and 1700 Å. The fit was limited to the wavelength intervals (1265:1285), (1305:1320), (1345:1370), (1415:1490), (1570:1590), and (1690:1710) to avoid stellar and interstellar absorption features following Calzetti et al. (1994).

The line flux of He II 1640 was taken as the sum of pixels in the 1D spectrum between 1631 and 1652 Å, and measured relative to the continuum as determined from the fit for β. The He II 4686 line is part of the "blue bump" which is a blend of nebular [Fe III] 4658 with multiple stellar lines, where the strengths and widths of those lines depend upon the types of W-R stars present (i.e., WN or WC, and depending on the mix of sub-types). Proper deblending can be a challenge, and requires high resolution spectra with high signal to noise. Here, we simply take the line flux as the sum of pixels in the 1D spectrum between 4670 – 4705 Å. The width of the window is based on the typical width of the observed He II 1640 line, which to first order is expected to be comparable to the one of the 4686 line (Hillier 1987; their Figures 5 and 6). The continuum is fit between 4000 – 5500 Å, excluding regions with known emission lines. We note that the STIS spectra are too low S/N to resolve the aforementioned blending of multiple lines in the "blue bump". The choice of the fitting window is therefore approximate and we expect an additional systematic error of ~30% on top of the random measurement errors for the line flux of He II 4686. For a more accurate measurement of the He II 4686 flux, the contribution from other lines to the "blue bump" can be constrained from models representing the expected properties of the star cluster region such as fraction of WN and WC stars, metallicity, or age (e.g., Sidoli et al. 2006). However, this is beyond the scope of this paper.

The Balmer-line fluxes are measured based on Gaussian-profile fits to the emission lines. The fluxes are consistent with those obtained as simple sums to within a few percent. There is no evidence of stellar absorption and no correction is applied.

For the conversion of Balmer and helium ratios and UV continuum slopes to attenuation values expressed in $E(B{-}V)$, we compare three reddening curves cited in Table 5: (1) LMC (Gordon et al. 2003); (2) $z \approx 2$ star forming galaxies (Reddy et al. 2015); (3) local starbursts (Calzetti et al. 2000). We note that the Reddy and Calzetti reddening curves are very similar and give virtually identical results. The LMC reddening curve is more appropriate for application to unresolved star clusters relative to the Calzetti curve (Sidoli et al. 2006). Hence we prefer the LMC curve for dereddening the Balmer- and helium-line ratios, but we list the results for both reddening laws in Table 6. We find minimal differences of less than 0.06 mag in *E(B–V)* if the correction is done with either curve. For the UV continuum slope, we applied Reddy et al. who provide a formulation of the relation between *E(B–V)* and β that is based on an intrinsic UV continuum slope of -2.44.

### *NGC 3049*

The STIS aperture is aligned along the bar of the galaxy NGC 3049. We extract the spectrum of the brightest component (component "B" in González Delgado et al. 2002) using extraction windows of 11 pixels for both the UV and optical spectrum. The other components are too faint in the optical to provide robust measurement of He II 4686 and the Balmer lines, hence we exclude them from our analysis. The broad UV and optical He II lines as well as the Balmer lines are well detected in cluster B. Strong interstellar [Fe II] and [Al II] absorption lines are detected close to He II 1640, as expected due to the high metallicity of this galaxy (1.2 $Z_{\odot}$). There is no evidence of narrow, nebular He II emission in the optical or UV spectra obtained with STIS. Guseva et al. (2000) reported faint nebular He II 4686 at a level that would not be detectable in the STIS spectrum. We measure $F(4686) = (3.4 \pm 0.3) \times 10^{-15}$ erg $s^{-1}$ $cm^{-2}$ for the optical He II line, which is consistent with the value obtained by González Delgado et al. (2002), who give a total flux of $1.1 \times 10^{-14}$ erg $s^{-1}$ $cm^{-2}$ for the 4640+4650+4686 bump combined. González Delgado et al. point out a flux difference of a factor of two with respect to Schaerer et al. (1999), which they ascribe to systematic offsets between space- and ground-based data. González Delgado et al. did not analyze the line flux of the 1640 line. Therefore no comparison with our value of *F*(1640) = $(9.8 \pm 1.8) \times 10^{-15}$ erg $s^{-1}$ $cm^{-2}$ can be made. Their derived value for the reddening from the UV continuum slope of *E(B-V)* = 0.2 agrees with ours (see the next paragraph).

We find total *E(B-V)* values (i.e., due to Galactic foreground dust plus intrinsic dust within NGC 3049) of $0.24^{+0.01}_{-0.01}$ (Reddy), $0.22^{+0.05}_{-0.04}$ (LMC), and $0.26^{+0.03}_{-0.04}$ (LMC) based on the UV continuum, the He II line ratio, and the Balmer decrement, respectively. The derived reddening values are consistent across all estimators within small error bars, suggesting very similar attenuation in nebular regions and stars. Reddening due to Milky Way foreground dust is *E(B-V)* = 0.036 based on maps of Schlegel et al. (1998), indicating a high internal reddening.

### *NGC 5398 (Tol 89)*

The STIS observations target the giant H II region commonly referred to as Tol 89 in the galaxy NGC 5398. We extracted five spectra corresponding to the five clusters studied by Sidoli et al. (2006), which they refer to as A1, A2, A3, A4, B1. Apertures of 11, 9, 11, 11, 11 pixels are used for the UV spectrum, and 11, 7, 11, 11, 11 pixels for the optical spectra, respectively. For A3 and B1, W-R He II lines are not detected in the UV or optical. Sidoli et al. detected narrow, nebular He II 4686 in cluster B1. The absence of W-R stars in this cluster implies the presence of ionization sources other than W-R stars for nebular He II. A prime mechanism for the production of nebular He II are high-energy photons originating from a population of massive X-ray binaries (Schaerer et al. 2019). For cluster A4, neither the optical He II line nor the UV continuum is detected (though He II 1640 is strong). Thus, we focus the analysis on clusters A1 and A2 only.

Cluster A1 is the brightest UV source among the clusters in the STIS aperture. The He II 1640 and Hα lines are strong and well detected. The "blue bump" and the Hβ line are weaker, as expected, but also detected. The optical continuum rises to the blue, but is faint and noisy due to the short exposure time (~800 s). The total *E(B–V)* values (i.e., due to Milky Way foreground dust plus dust internal to NGC 5398) are $0.16^{+0.20}_{-0.20}$ (Reddy), $0.18^{+0.09}_{-0.06}$ (LMC), and $0.39^{+0.34}_{-0.25}$ (LMC) based on the UV continuum, the He II line ratio, and the Balmer-line ratio, respectively.

Cluster A2 is the second brightest UV source in the aperture and is separated from A1 only by a few tenths of an arcsec. A1 and A2 are resolved in the UV spectrum but barely resolved in the optical spectrum. A2's He II 1640 flux is comparable to that of A1, but its He II 4686 flux is much weaker. The *E(B–V)* values are $0.05^{+0.03}_{-0.04}$ (Reddy), $0.06^{+0.06}_{-0.06}$ (LMC), and $0.05^{+0.47}_{-0.42}$ (LMC) based on the UV continuum, the He II line ratio, and the Balmer-line ratio, respectively.

Thus, for both A1 and A2 the dust attenuation inferred from the He II lines and the UV continuum are consistent. However, we cannot rule out factor of two differences relative to the Balmer reddening and a possible a dependence of reddening with age because of the uncertainties and the low overall reddening in these objects. Reddening due to Milky Way foreground dust is *E(B–V)* = 0.066 based on maps of Schlegel et al. (1998), so the reddening due to dust within the galaxy is quite low. Studies of local star-forming galaxies suggest higher reddening corrections for the gas than for the stars (Calzetti 2001). Taken at face value, the higher Balmer-line reddening in Tol 89 agrees with these results. However, the reddening is not well constrained by the Balmer decrement; within the large errors, the attenuation from the H$\alpha$/H$\beta$ ratio is consistent with the values determined from the other two methods.

These results can be compared with those of Sidoli et al. (2006). These authors present Very Large Telescope (VLT) high spectral resolution, high signal-to-noise UV–Visual Echelle Spectrograph (UVES) data, and also analyze the STIS data for the purpose of characterizing the physical properties of clusters, with focus on the W-R populations in Tol 89. The UVES spectra provide the high quality measurements of the optical lines required for a more robust analysis of the "blue bump" and measurement of the He II 4686 line. They were obtained with a slit width of 1.4". Clusters A1 and A2 are unresolved. An intrinsic He II 4686 flux of $(3.9 \pm 0.4) \times 10^{-15}$ erg s$^{-1}$ cm$^{-2}$ (observed as $(2.85 \pm 0.3) \times 10^{-15}$ erg s$^{-1}$ cm$^{-2}$ assuming $E(B-V)$ = 0.07) is reported for A1+A2. This is 37% larger than the observed total of $2.07^{+0.20}_{-0.22} \times 10^{-15}$ erg s$^{-1}$ cm$^{-2}$ measured here. The He II 4686 flux of Sidoli et al. used in combination with the He II 1640 flux for A1+A2 reported in Table 6 yields $E(B-V)$ = $0.21 \pm 0.4$.

## 7. Discussion and Conclusions

The technique of utilizing the strengths of the UV and optical W-R He II lines to infer dust reddening in star-forming galaxies is a novel alternative to other established methods. A plethora of literature values of optical W-R line fluxes in star-forming galaxies already exists. This method can also be applied to active galactic nuclei, where both transitions are readily detected as broad nebular emission lines. Together with photoionization models, the observed line ratios can serve as an indicator of dust attenuation similar to the case of star-forming galaxies (Osterbrock 1989; Netzer 1990).

We remind the reader that our definition of the optical W-R "bump" refers to the He II 4686 line only. W-R galaxies also exhibit the N III/C III bump at 4640/50 Å. N III itself is a blend of N III 4634 and 4640 originating in late WN stars. The N III lines gradually weaken towards earlier WN stars, in which N V 4604 and 4620 is seen instead (Hainich et al. 2014). At low spectral resolution, He II 4686 and N III/C III 4640/50 can be blended together, and some authors (e.g., López-Sánchez & Esteban 2010) define the W-R bump as the combined N III/C III + He II feature. In the present work we adopt the commonly used definition for the W-R bump as due to He II 4686 alone (e.g., Vacca & Conti 1992). Sufficiently high spectral resolution is also needed to identify and correct for the contamination of the broad He II 4686 line by narrow nebular emission of [Fe III] 4658 and [Ar IV] 4711 (see Figure 7 of Schaerer et al. 1999), even though both lines are often quite weak. Finally, a subset of W-R galaxies in the local universe shows narrow, nebular He II 4686 on top of the broad, stellar He II 4686 (Shirazi & Brinchmann 2012). Current spectroscopic surveys in the local universe (e.g., the Sloan Digital Sky Survey; York et al. 2000) and near cosmic noon (e.g., the MOSFIRE Deep Evolution Field; Kriek et al. 2015) have a typical resolving power of at least $R = 10^3$. HST's spectrographs match these resolutions as well. Therefore nebular contributions to the broad W-R lines can be recognized and corrected for. There is no unique correlation between the presence of the nebular and the stellar He II line, making it unlikely for W-R stars to be the ionizing source of the nebular lines. Alternatively, massive X-ray binaries as powering sources have been proposed by Schaerer et al. (2019). Gräfener & Vink (2015) considered yet another origin of narrow He II: very massive stars can form in a low-metallicity environment having 1% of the solar abundance. Such stars have strong winds with low velocities due to their proximity to the Eddington instability limit. Such winds can mimic a narrow, stellar line. Star-forming galaxies in this metallicity regime have not yet been discovered but are prime objects for future study with the James Webb Space Telescope (JWST), which will extend the accessible redshift of the He II 4686 line beyond its current ground-based limit of $z \approx 4$.

W-R stars come in two flavors, nitrogen-rich WN stars and carbon-rich WC stars. We consider only WN stars as tracers of dust attenuation. In order to apply the proposed technique to the integrated spectra of star-forming galaxies, we need to evaluate the contribution of WC stars, which are detected in galaxy spectra via their strong C IV 5801 and 5812 lines, as well as in the C III 4650 blend close to the He II 4686 line. Would WC stars make a noticeable contribution to He II 1640 and 4686? Individual WC stars do of course show He II 1640 and 4686 but the He II lines are weaker than the prominent carbon lines of C IV 1550 and C III 1908 in WC stars of any spectral type (Sander et al. 2012). However, a broad C III 1908 bump is not observed in star-forming galaxies, which places a strong constraint on the contribution of WC stars to the He II lines. This empirical result is the consequence of the lower WC luminosities compared with WN stars (Sander et al., their Figure 9) as well as their shorter lifetimes. The different lifetimes of WN and WC stars are reflected in the relative numbers of WN and WC stars in the LMC, where the census is complete (Massey et al. 2014, 2015, 2017). The final survey of Neugent et al. (2018) gives 154 W-R stars, out of which only 22 are classified as WC. The WC/WN ratios determined for other Local Group galaxies suggest a correlation with oxygen abundance, ranging between 0.6 in M31 and 0.1 in the SMC (Massey 2013). We conclude that our restriction to WN stars as the source of the He II 1640 and 4686 lines in galaxy spectra is justified.

Throughout this paper we assumed W-R stars to be the evolved counterparts of massive O stars. We emphasize that our conclusions do not depend on this assumption. W-R stars can have any origin or formation channel, e.g., via single-star evolution (Meynet et al. 2017) or as the product of interaction in binary stars (Eldridge et al. 2017). The ratio of the two He II lines would not be affected by the particulars of the formation and evolution of the underlying stars. We also stress that the line ratio is not affected by IMF variations in the stellar population, nor does it depend on the details of the star-formation history. Both lines are emitted in the winds of the same stellar type. If changes of the IMF favor one type over the other, we would still observe the same line ratio as we found no indication of a dependence of the ratio on W-R sub-type, except for types later than WN8, whose line luminosities are low and therefore do not make a significant contribution. We confirmed this behavior by performing a series of population synthesis calculations with the Starburst99 code (Leitherer et al. 1999; 2014), which result in a near-constant ratio of the 1640 Å over the 4686 Å W-R line for a broad range of IMF exponents. An instantaneous starburst model of age 4 Myr using rotating single-star evolutionary tracks with solar chemical composition predicts $R_0$ = 7.41, 7.76, and 7.08 for IMF exponents of 2.3 (Salpeter), 1.5, and 3.0, respectively.

When applied to star-forming galaxies at low redshift, space-UV spectroscopy is required. This makes this method relatively resource intensive, as HST data are needed. However, we do not anticipate the ratio of the UV and optical He II lines to be used as extensively as the traditional Balmer-line ratio for reddening studies. Rather, we see the new method as a targeted approach to answer specific questions in a well-defined sample of galaxies: How do reddening determinations in galaxies from this and other methods compare? Is the dust attenuation of the gas, as measured from the Balmer decrement, different from that of the contemporaneous generation of stars, as measured by the W-R lines? Does the dust reddening of the stars change with age from the youngest population, as determined from the W-R line ratio, to the older B-star population measured from the UV spectral slope? Answering these questions requires panchromatic UV-to-optical spectra of star-forming galaxies obtained at *co-spatial* location through *matching apertures*. In this paper we presented an initial application of the new technique to existing archival spectra of two young starburst clusters. The data were not taken through matching apertures but serve as a proof of concept to guide future observations.

*Acknowledgments.* Useful email correspondence with Paul Crowther and Andreas Sander are gratefully acknowledged. We are grateful to Daniela Calzetti and Karl Gordon for their comments on the reddening law section of the paper. Support for this work has been provided by NASA through grant number AR-15036 from the Space Telescope Science Institute, which is operated by AURA, Inc., under NASA contract NAS5-26555. This research has made extensive use of the NASA/IPAC Extragalactic Database (NED) which is operated by the Jet Propulsion Laboratory, California Institute of Technology, under contract with the National Aeronautics and Space Administration.

# Figures

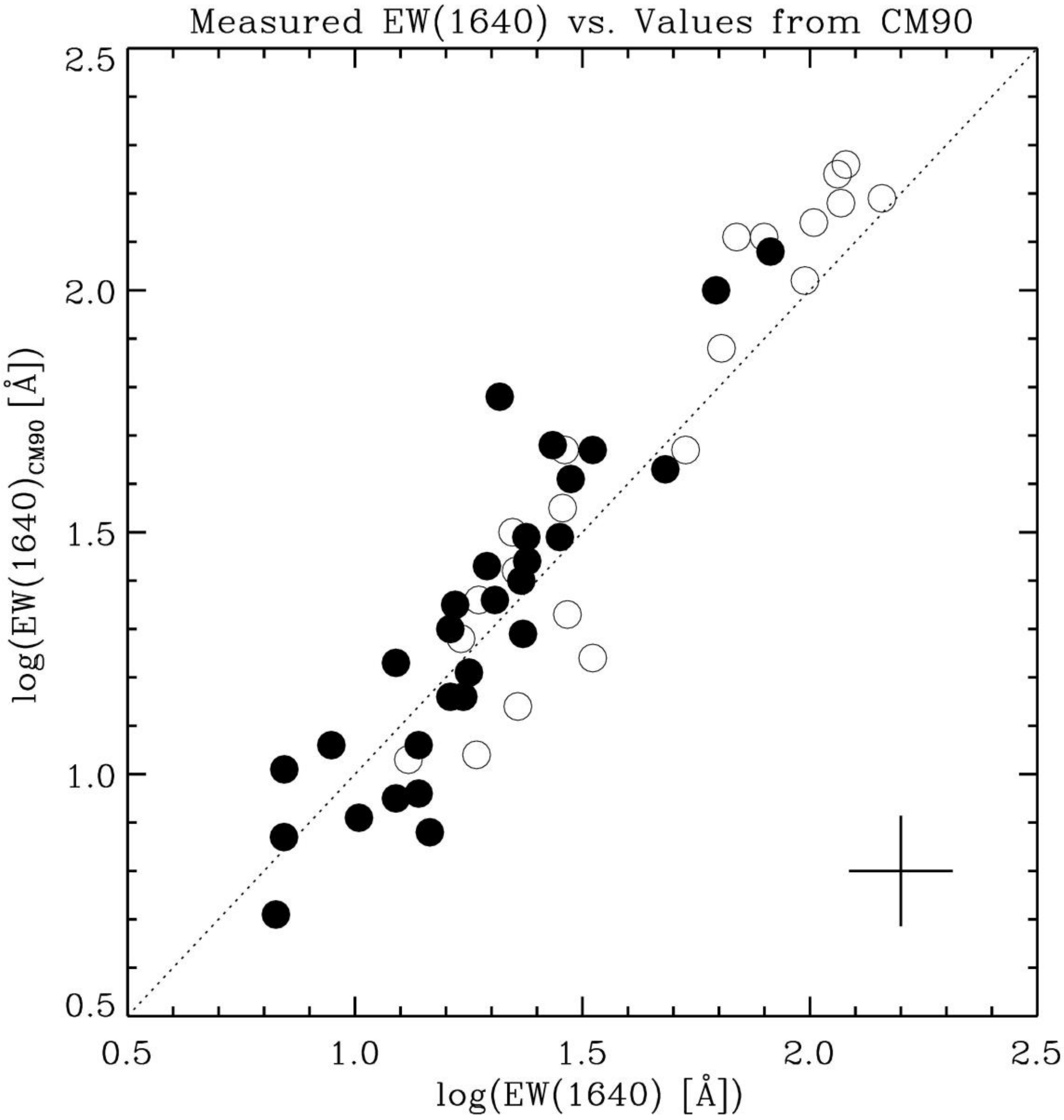

Figure 1. Comparison of the equivalent width of He II 1640 measured in this work (abscissa) and the values published by Conti & Morris (1990; ordinate). Solid symbols: Milky Way stars; open symbols: LMC stars; dotted line: one-to-one relation. The cross indicates the error of the individual measurements. The dotted line is the one-to-one relation.

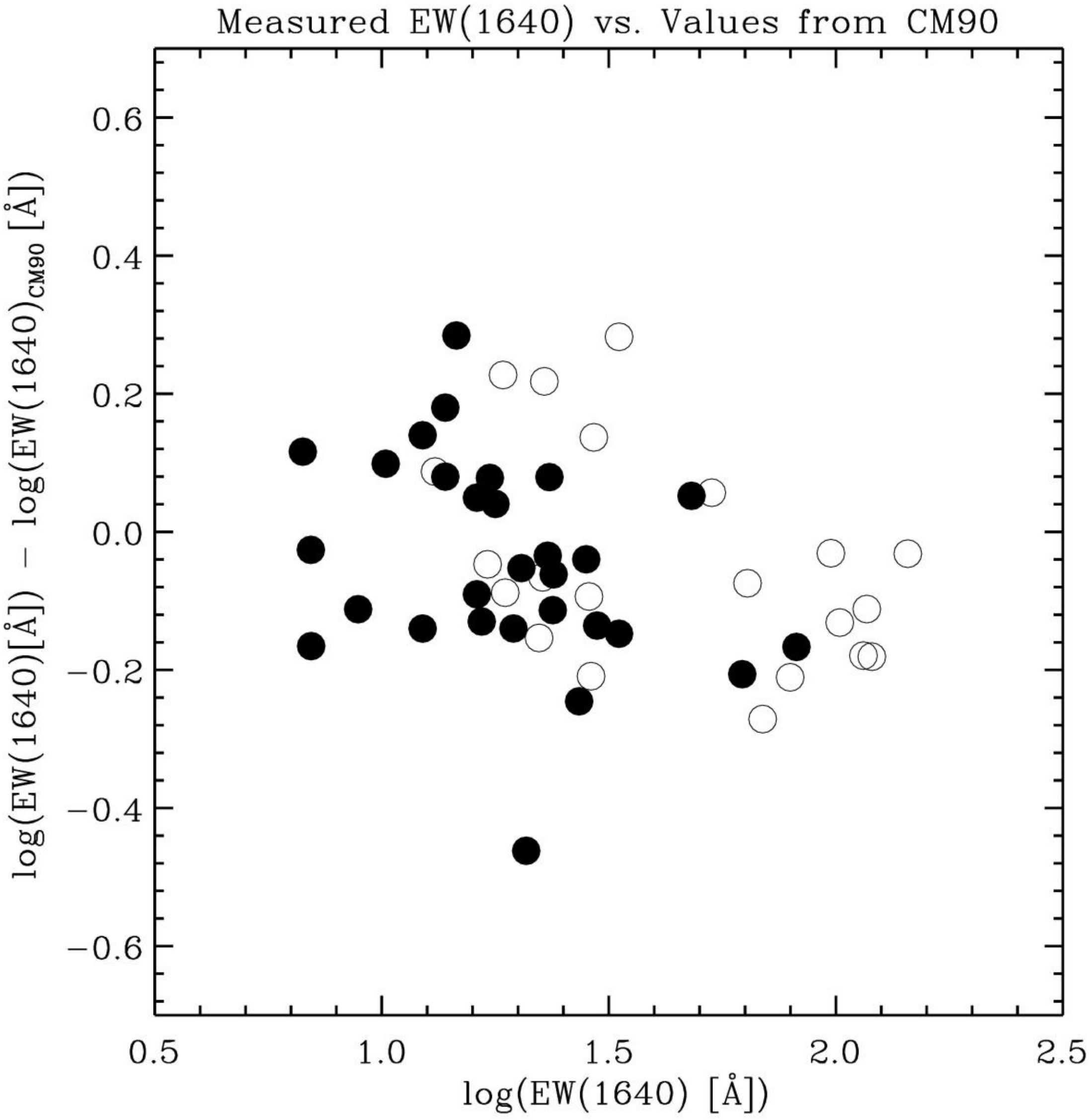

Figure 2. Comparison of the logarithmic ratio of the equivalent width of He II 1640 measured in this work over the values published by Conti & Morris (1990; ordinate) versus the values in this work (abscissa). The symbols are as in Figure 1.

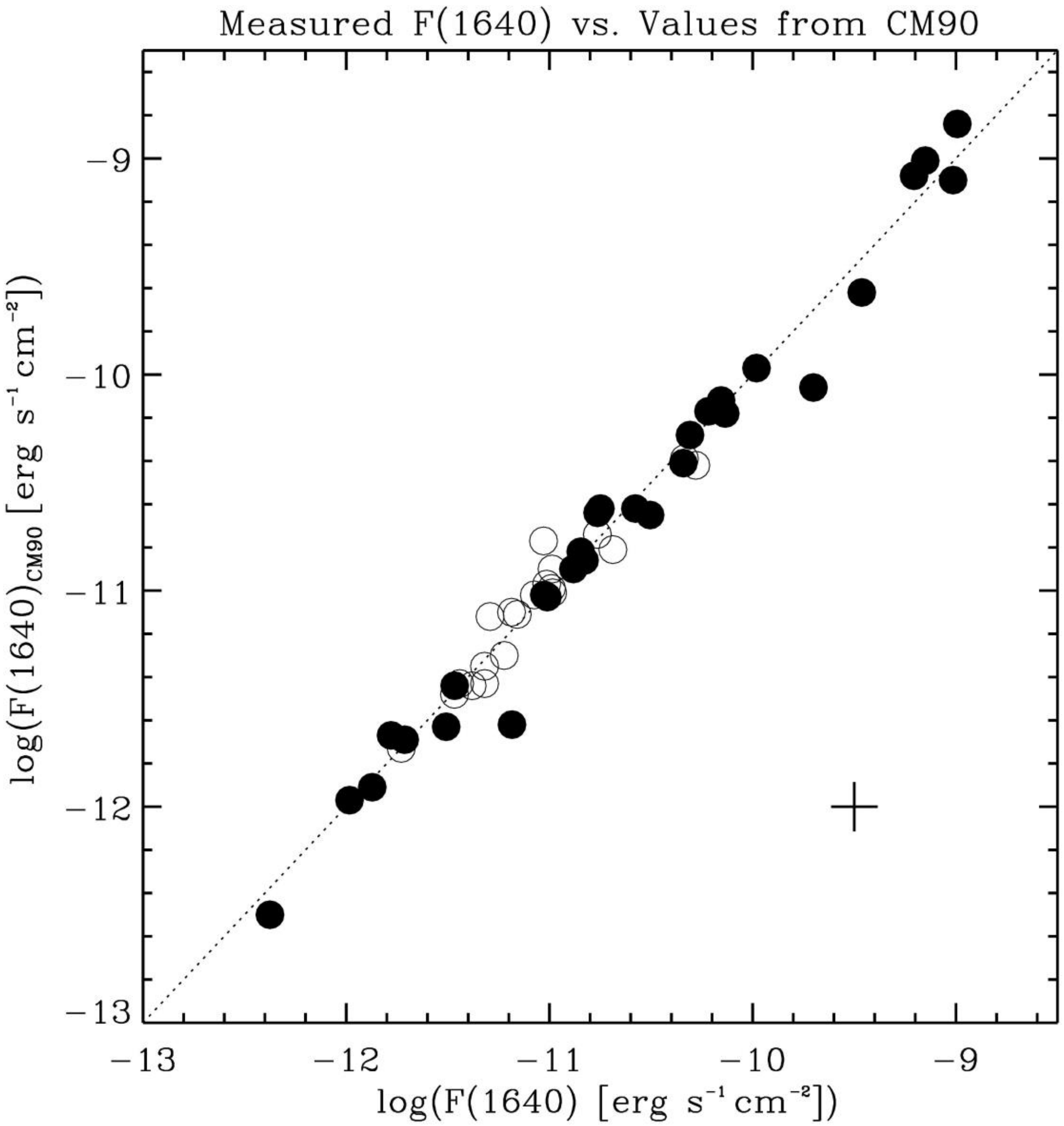

Figure 3. Same as Figure 1, but for the line flux measurements.

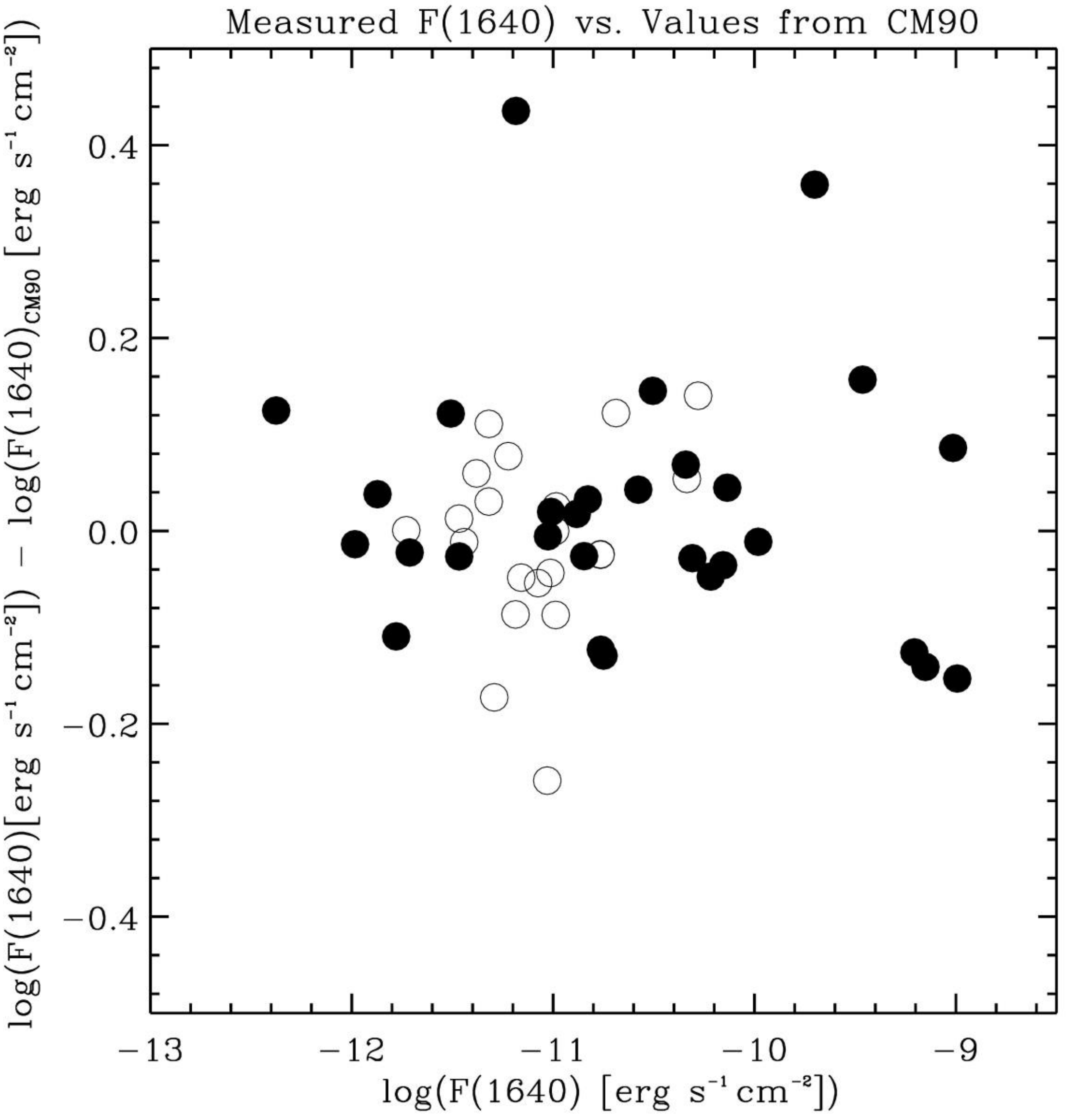

Figure 4. Same as Figure 2, but for the logarithmic ratio of the line fluxes.

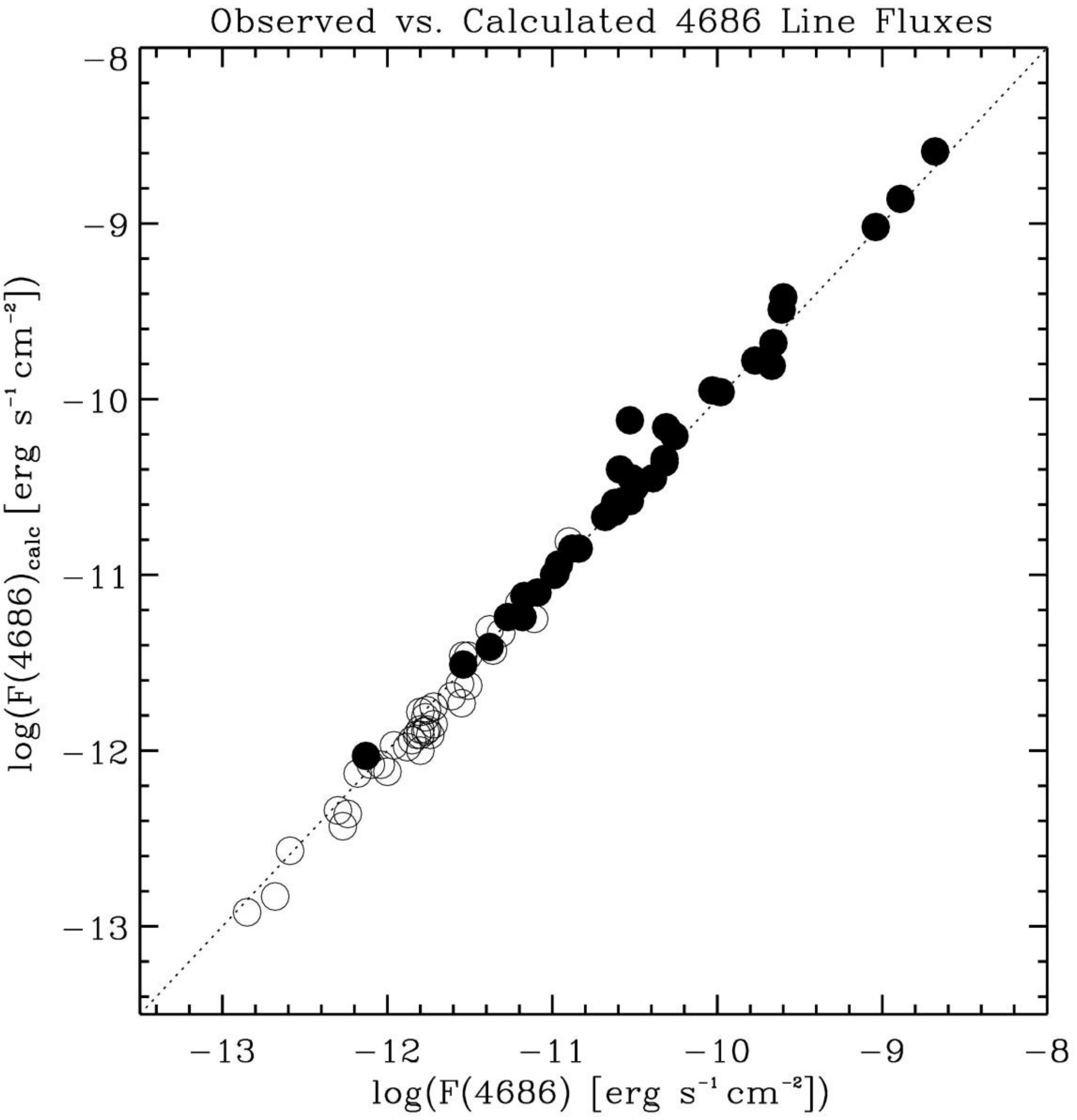

Figure 5. Comparison of the observed He II 4686 line fluxes (abscissa) and the values calculated from the He II 4686 equivalent width and *b*, *v* photometry (ordinate). Solid symbols: Milky Way stars; open symbols: LMC stars; dotted line: one-to-one relation.

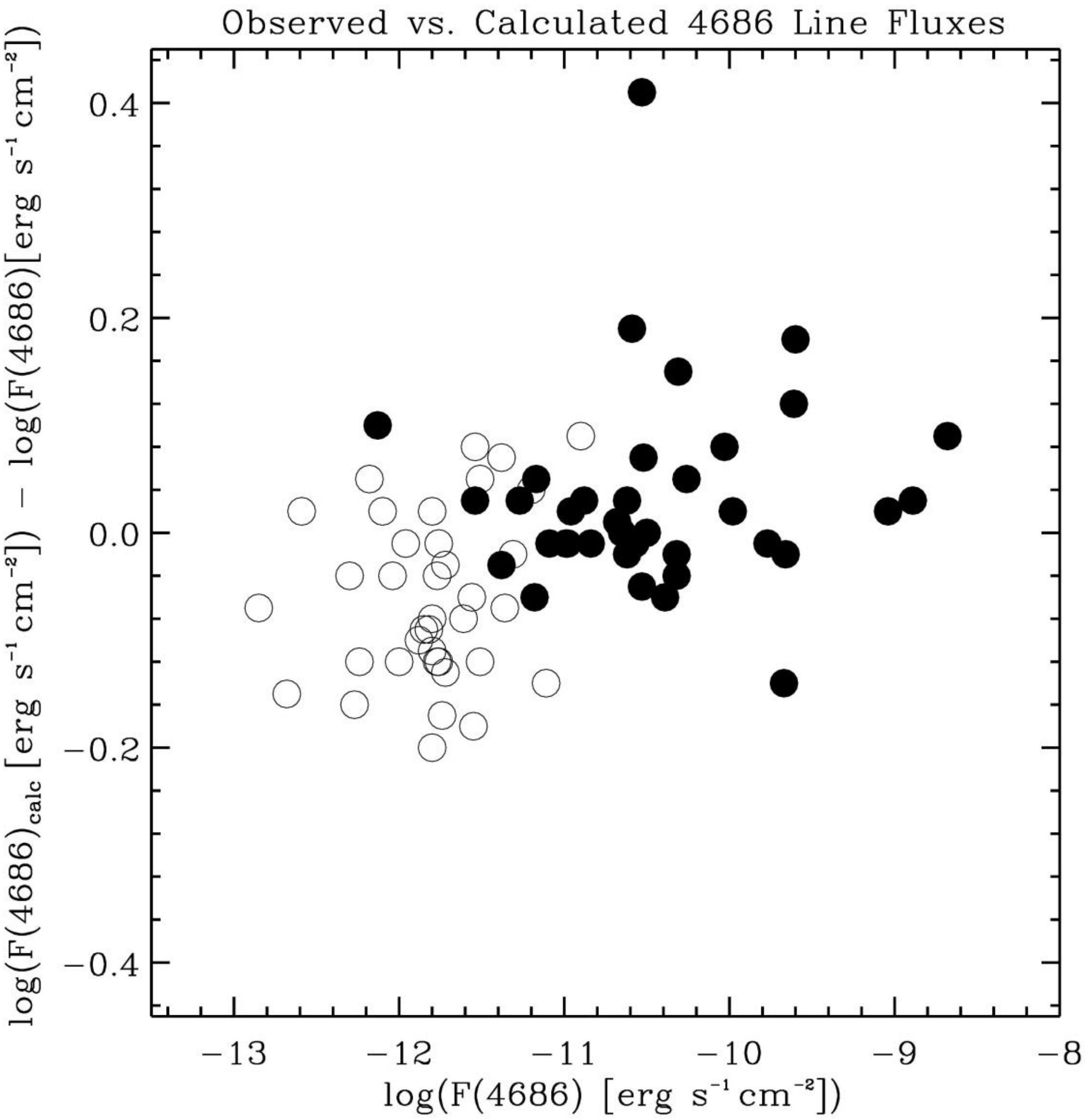

Figure 6. Same as Figure 5, but for the logarithmic ratio of the calculated values of the He II 4686 fluxes over the observed values versus the observed values.

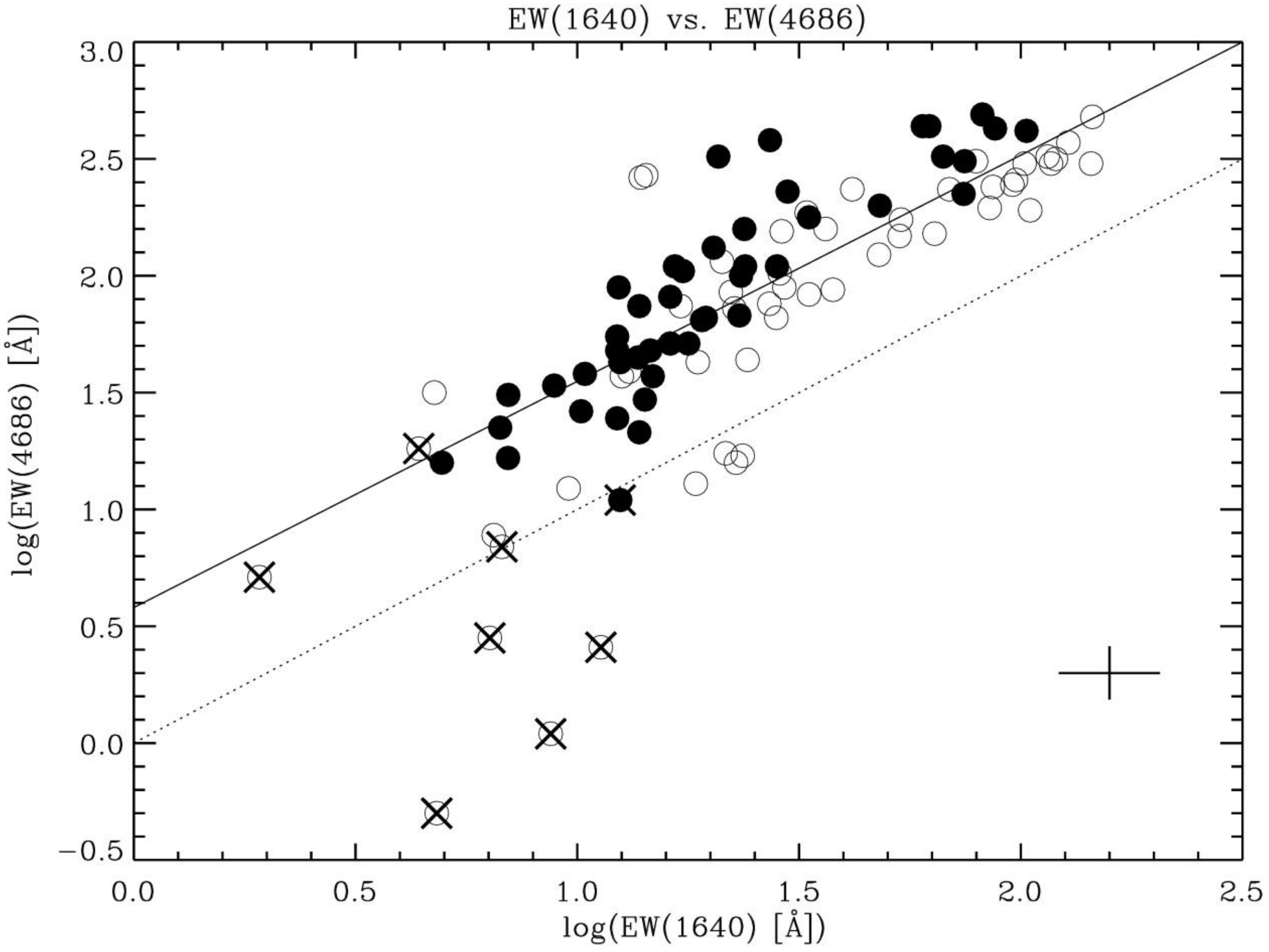

Figure 7. Comparison of the observed He II 1640 (abscissa) and He II 4686 equivalent widths (ordinate). Solid symbols: Milky Way stars; open symbols: LMC stars. W-R stars of spectral types WN9 – 11 are marked with a cross. Dotted line: one-to-one relation; solid line: least-square fit to the data excluding the WN9 – 11 stars. The cross at the lower right indicates the error of the individual measurements.

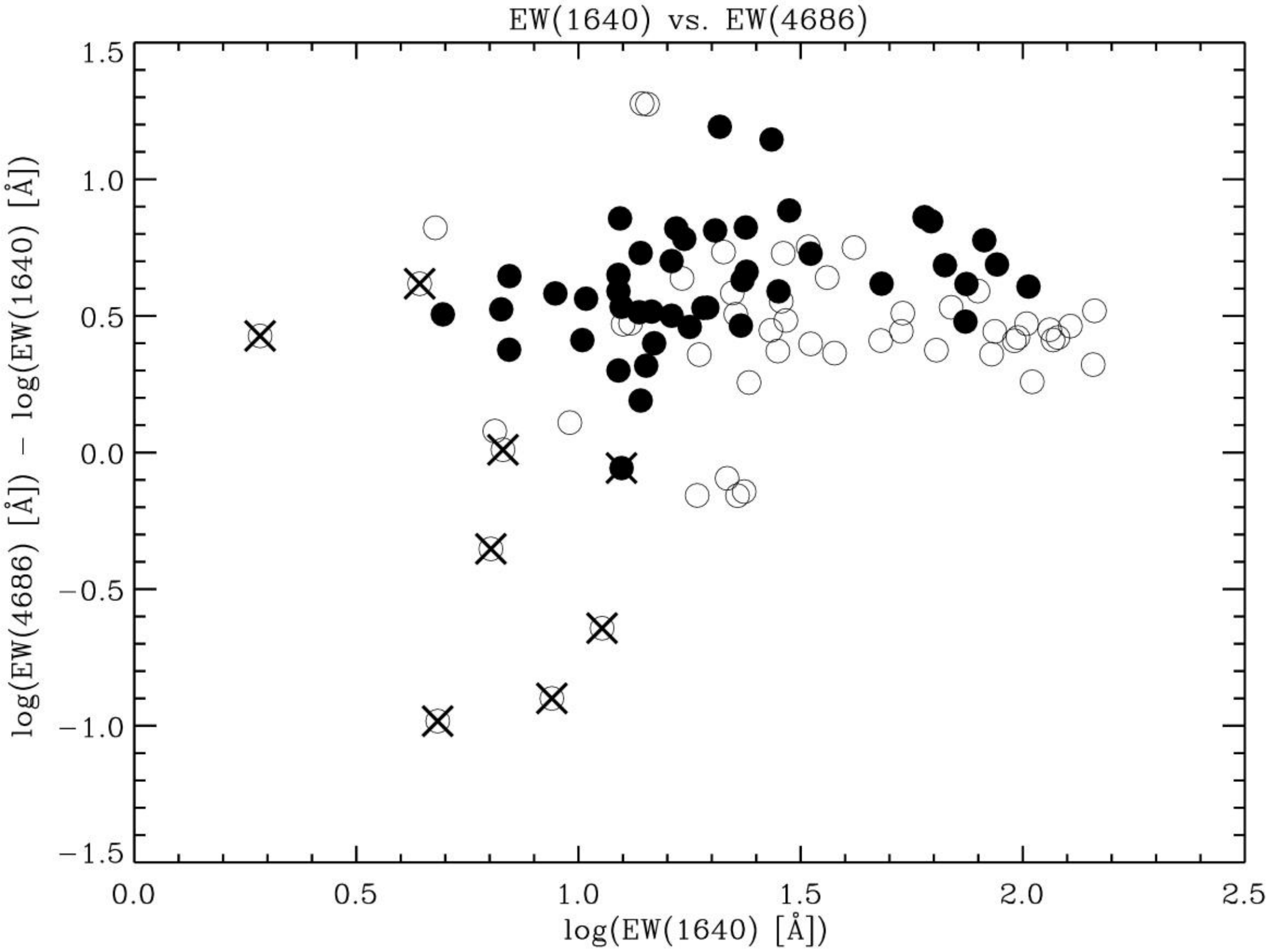

Figure 8. Same as Figure 7, but for the logarithmic ratio of the values of the equivalent widths of He II 4686 over He II 1640 versus the values of He II 1640.

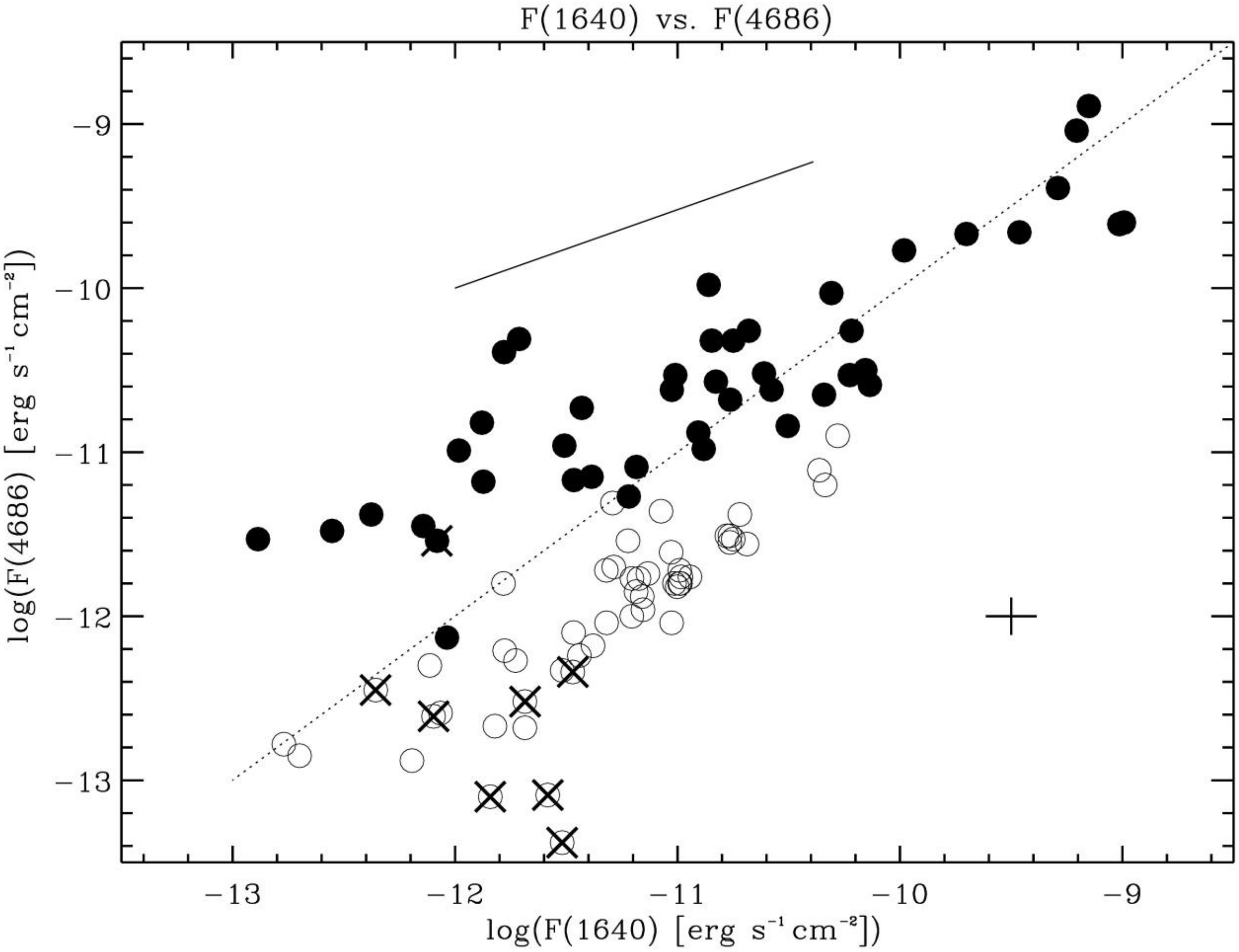

Figure 9. Same as Figure 7, but for the line flux measurements. The solid line is the reddening vector for *E(b–v)* = 0.5.

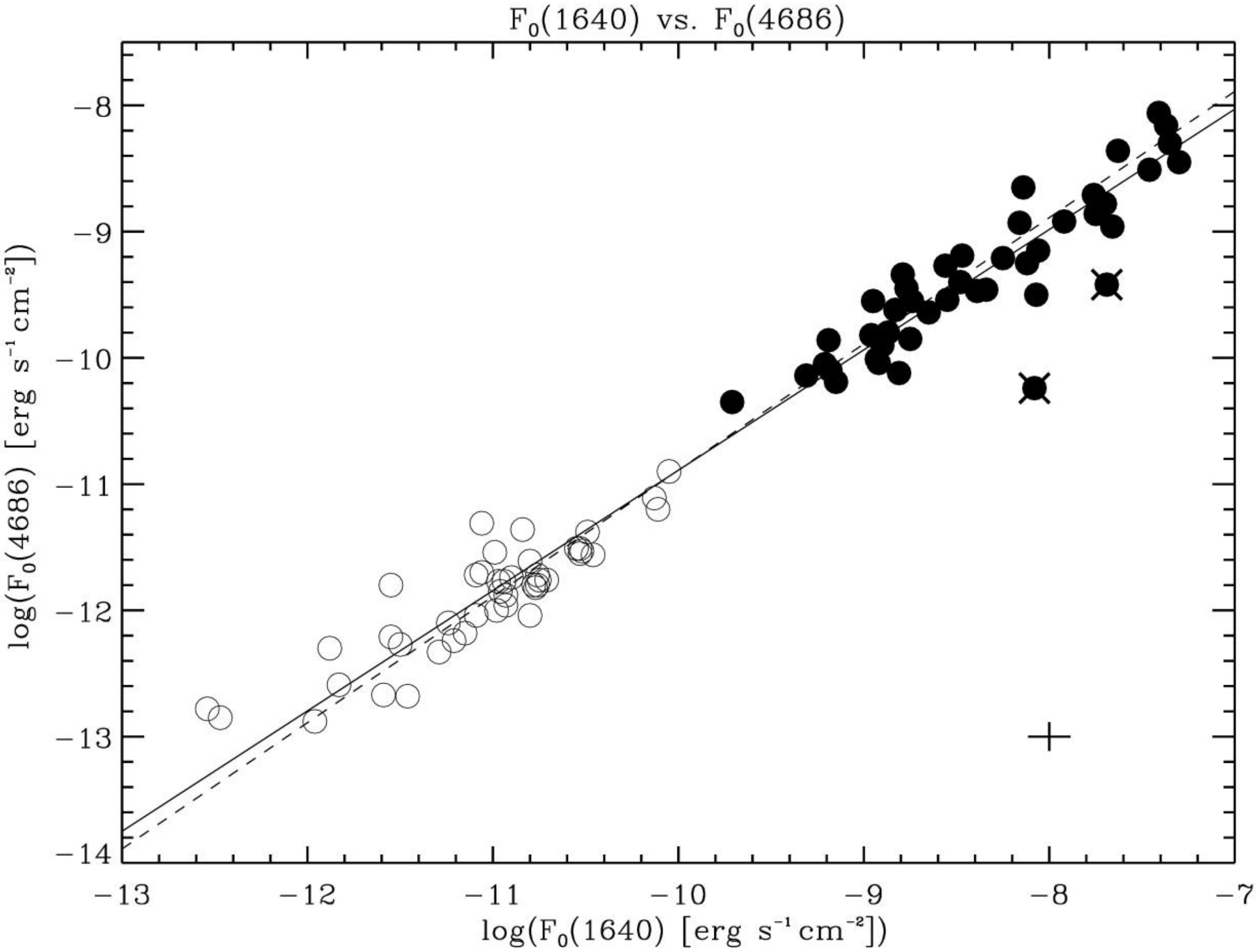

Figure 10. Relation between the dereddened He II 1640 (abscissa) and He II 4686 line fluxes (ordinate). Solid symbols: Milky Way stars; open symbols: LMC stars. WR 25 and WR 43a are marked with a cross. Dashed line: relation for a constant ratio of $R_0$ =7.76; solid line: least-square fit to the data excluding the two marked stars. The cross at the lower right indicates the error of the individual measurements.

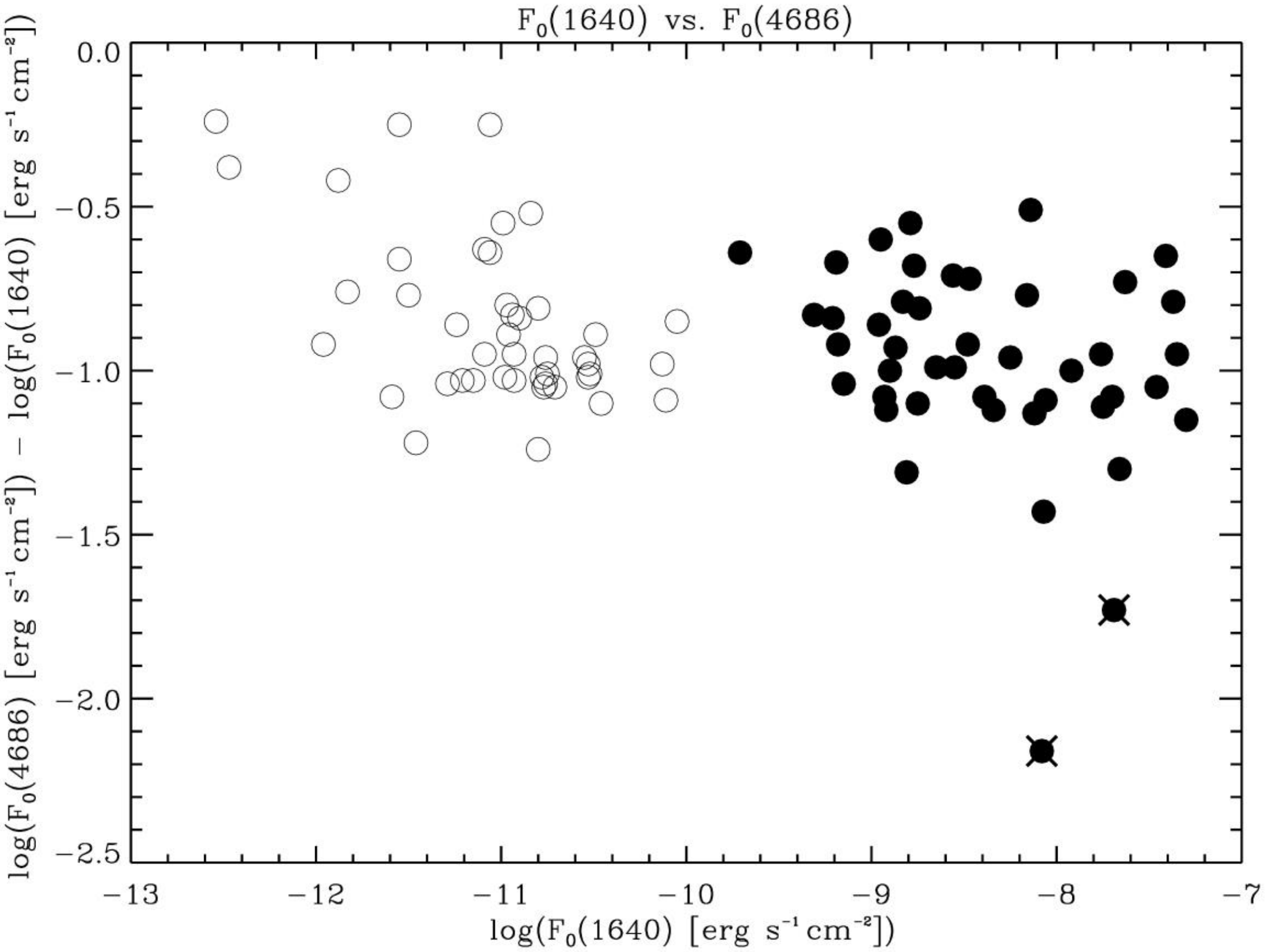

Figure 11. Same as Figure 10, but for the logarithmic ratio of the values of the dereddened He II 4686 over He II 1640 fluxes versus the values of the He II 1640 fluxes.

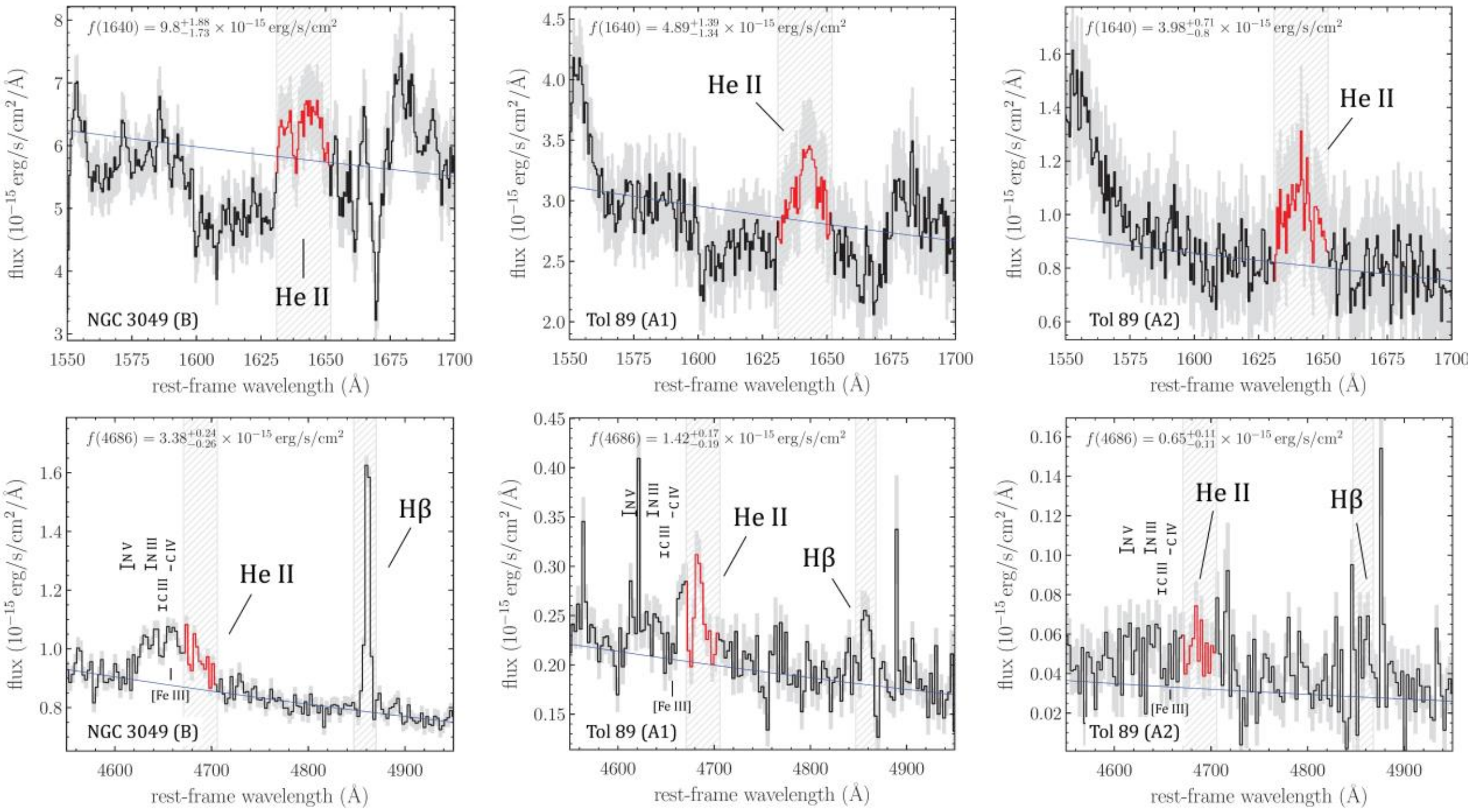

Figure 12. Rest-frame spectral regions around the He II lines at 1640 (top) and 4686 Å (bottom) for the two W-R sources NGC 3049 (left) and Tol 89 (center and right). The black line shows the observations with measurement uncertainties (gray). The extraction windows are indicated by the hatched regions and the summed spectral region is drawn in red. The continuum fit is shown in blue. Other stellar and nebular emission lines are also indicated.

# Tables

Table 1. Milky Way program stars[2].

| W-R ID[a] | Other ID[b] | R.A.[c] (h m s) | Dec.[d] (° ′ ″) | Spectral Type[e] | IUE ID[f] | log(*EW*(1640))[g] (Å) | log(*F*(1640))[h] ($erg\ s^{-1}\ cm^{-2}$) |
|---|---|---|---|---|---|---|---|
| WR 1 | HD 4004 | 00 43 28.40 | +64 45 35.4 | WN4 | SWP13911 | 1.91 | -10.31 |
| WR 2 | HD 6327 | 01 05 23.03 | +60 25 18.9 | WN2 | SWP20303 | 1.68 | -10.58 |
| WR 3 | HD 9974 | 01 38 55.63 | +58 09 22.7 | WN3+O4 | SWP20293 | 1.09 | -10.50 |
| WR 6 | HD 50896 | 06 54 13.05 | -23 55 42.1 | WN4 | SWP20901 | 1.87 | -7.78 |
| WR 7 | HD 56925 | 07 18 29.13 | -13 13 01.5 | WN4 | SWP03541 | 1.94 | -10.61 |
| WR 8 | HD 62910 | 07 44 58.22 | -31 54 29.6 | WN7/WCE+? | SWP10747 | 1.22 | -11.01 |
| WR 10 | HD 65865 | 07 59 46.25 | -28 44 03.1 | WN5ha | SWP13912 | 1.25 | -11.18 |
| WR 12 | CD -45 4482 | 08 44 47.25 | -45 58 55.5 | WN8h+? | SWP18473 | 1.14 | -11.51 |
| WR 16 | HD 86161 | 09 54 52.91 | -57 43 38.3 | WN8h | SWP06970 | 1.14 | -10.23 |
| WR 18 | HD 89358 | 10 17 02.28 | -57 54 46.9 | WN4 | SWP38381 | 2.01 | -10.68 |
| WR 21 | HD 90657 | 10 26 31.42 | -58 38 26.2 | WN5+O4-6 | SWP34767 | 1.21 | -10.83 |
| WR 22 | HD 92740 | 10 41 17.52 | -59 40 36.9 | WN7h+O9 | SWP16083 | 1.01 | -9.01 |
| WR 24 | HD 93131 | 10 43 52.27 | -60 07 04.0 | WN6ha | SWP01634 | 0.84 | -8.99 |
| WR 25 | HD 93162 | 10 44 10.39 | -59 43 11.0 | WN6h+O4f | SWP49730 | 0.84 | -10.14 |
| WR 31 | HD 94546 | 10 53 44.83 | -59 30 46.7 | WN4+O8 | SWP17066 | 1.28 | -10.91 |
| WR 40 | HD 96548 | 11 06 17.21 | -65 30 35.2 | WN8h | SWP05826 | 1.16 | -9.70 |
| WR 43a | HD 97950-A1 | 11 15 07.47 | -61 15 38.2 | WN6ha | SWP22528 | 0.69 | -12.04 |
| WR 44 | LS 2289 | 11 16 57.86 | -59 26 24.0 | WN4+OB? | SWP38380 | 1.87 | -11.22 |
| WR 46 | HD 104994 | 12 05 18.73 | -62 03 10.1 | WN3p+OB? | SWP18476 | 1.24 | -10.34 |
| WR 47 | HDE 311884 | 12 43 51.01 | -63 05 14.8 | WN6+O5 | SWP45171 | 1.21 | -11.98 |
| WR 55 | HD 117688 | 13 33 30.13 | -62 19 01.2 | WN7 | SWP13896 | 1.31 | -11.03 |
| WR 58 | LS 3162 | 13 49 04.52 | -65 41 56.0 | WN4/WCE | SWP41774 | 1.82 | -11.39 |
| WR 61 | LS 3208 | 14 13 03.53 | -65 26 52.7 | WN5 | SWP09188 | 1.52 | -11.47 |
| WR 71 | HD 143414 | 16 03 49.35 | -62 41 35.8 | WN6+OB? | SWP09761 | 1.38 | -10.22 |
| WR 75 | HD 147419 | 16 24 26.23 | -51 32 06.1 | WN6 | SWP11124 | 1.32 | -11.78 |
| WR 78 | HD 151932 | 16 52 19.25 | -41 51 16.2 | WN7h | SWP04335 | 1.10 | -9.29 |
| WR 85 | HD 155603-B | 17 14 27.13 | -39 45 47.0 | WN6h+OB | SWP45274 | 1.09 | -11.43 |
| WR 108 | HDE 313846 | 18 05 25.74 | -23 00 20.3 | WN9h+OB | SWP38443 | 1.10 | -12.08 |
| WR 110 | HD 165688 | 18 07 56.96 | -19 23 56.8 | WN5-6 | SWP38442 | 1.78 | -10.86 |
| WR 123 | HD 177230 | 19 03 59.02 | -04 19 01.9 | WN8 | SWP14716 | 1.15 | -12.14 |
| WR 127 | HD 186943 | 19 46 15.94 | +28 16 19.1 | WN3+O9.5 | SWP24646 | 1.29 | -10.76 |
| WR 128 | HD 187282 | 19 48 32.20 | +18 12 03.7 | WN4h+OB? | SWP06998 | 1.38 | -10.16 |
| WR 133 | HD 190918 | 20 05 57.33 | +35 47 18.2 | WN5+O9 | SWP17080 | 1.14 | -9.46 |
| WR 134 | HD 191765 | 20 10 14.20 | +36 10 35.1 | WN6 | SWP07161 | 1.79 | -9.21 |
| WR 136 | HD 192163 | 20 12 06.55 | +38 21 17.8 | WN6(h) | SWP04309 | 1.43 | -9.15 |

[2] The coordinates are from van der Hucht (2001) and may be different from the MAST coordinates.

| | | | | | | | |
|---|---|---|---|---|---|---|---|
| WR 138 | HD 193077 | 20 17 00.03 | +37 25 23.8 | WN5+B? | SWP15614 | 1.37 | -9.98 |
| WR 141 | HD 193928 | 20 21 31.73 | +36 55 12.8 | WN5+O5 | SWP16080 | 1.47 | -11.71 |
| WR 148 | HD 197406 | 20 41 21.55 | +52 35 15.2 | WN8h+B3 | SWP14713 | 0.83 | -11.87 |
| WR 151 | CX Cep | 22 09 33.47 | +57 44 30.7 | WN4+O5 | SWP28883 | 1.37 | -12.38 |
| WR 152 | HD 211564 | 22 16 24.05 | +55 37 37.2 | WN3h | SWP18450 | 1.45 | -10.88 |
| WR 153 | HD 211853 | 22 18 45.61 | +56 07 33.9 | WN6/WCE+O6 | SWP24981 | 1.09 | -10.75 |
| WR 155 | HD 214419 | 22 36 53.96 | +56 54 21.0 | WN6+O9 | SWP14170 | 0.95 | -10.85 |
| WR 156 | ALS 12730 | 23 00 10.13 | +60 55 38.4 | WN8h+OB? | SWP18465 | 1.09 | -12.89 |
| WR 157 | HD 219460 | 23 15 12.41 | +60 27 01.9 | WN5(+B1) | SWP17078 | 1.17 | -11.88 |
| WR 158 | ALS 13110 | 23 43 30.60 | +61 55 48.1 | WN7h+Be? | SWP39432 | 1.02 | -12.56 |

a. W-R identifier from van der Hucht (2001)
b. alternate designation
c. Right Ascension (2000) from van der Hucht (2001)
d. Declination (2000) from van der Hucht (2001)
e. spectral type from van der Hucht (2001)
f. identifier of the IUE spectrum
g. measured equivalent width of He II 1640
h. measured line flux of He II 1640

Table 2. LMC program stars[3].

| W-R ID[a] | Other ID[b] | R.A.[c] (h m s) | Dec.[d] (° ′ ″) | Spectral Type[e] | IUE ID[f] | log(*EW*(1640))[g] (Å) | log(*F*(1640))[h] (erg $s^{-1}$ $cm^{-2}$) |
|---|---|---|---|---|---|---|---|
| BAT99 1 | Brey 1 | 04 45 32.07 | -70 15 11.3 | WN3b | SWP33444 | 2.16 | -11.38 |
| BAT99 2 | Brey 2 | 04 49 36.22 | -69 20 54.2 | WN1 | SWP45505 | 1.56 | -12.06 |
| BAT99 3 | Brey 3 | 04 52 57.43 | -66 41 12.9 | WN4b | SWP33446 | 1.99 | -10.98 |
| BAT99 7 | HD 32109 | 04 55 31.29 | -67 30 01.7 | WN4b | SWP38449 | 2.16 | -10.36 |
| BAT99 12 | Brey 10a | 04 57 27.35 | -67 39 02.2 | WN6/O3f | SWP48107 | 0.98 | -11.82 |
| BAT99 13 | SK -66 40 | 04 57 41.05 | -66 32 41.9 | WN10 | SWP43589 | 1.05 | -11.84 |
| BAT99 14 | Brey 11 | 04 58 56.37 | -68 48 04.3 | WN4+OB? | SWP41715 | 1.58 | -11.03 |
| BAT99 15 | HD 268847 | 04 59 51.60 | -67 56 53.6 | WN4b | SWP09168 | 2.06 | -10.76 |
| BAT99 16 | HD 33133 | 05 03 08.84 | -66 40 57.3 | WN8h | SWP04364 | 1.35 | -10.76 |
| BAT99 17 | HD 269015 | 05 04 12.35 | -70 03 54.8 | WN4o | SWP04365 | 1.81 | -10.99 |
| BAT99 18 | Brey 15 | 05 05 08.34 | -70 22 44.7 | WN4h | SWP38236 | 1.68 | -11.21 |
| BAT99 19 | HD 34187 | 05 09 40.45 | -68 53 24.3 | WN4b+OB? | SWP38232 | 2.02 | -10.75 |
| BAT99 22 | HD 269227 | 05 13 54.40 | -69 31 46.1 | WN9h | SWP10729 | 0.83 | -11.47 |

[3] The coordinates are from Breysacher et al. (1999) and may be different from the MAST coordinates.

| | | | | | | | |
|---|---|---|---|---|---|---|---|
| BAT99 24 | HD 34783 | 05 14 12.83 | -69 19 25.5 | WN4b | SWP33448 | 2.11 | -10.78 |
| BAT99 26 | Brey 20 | 05 16 38.91 | -69 16 40.4 | WN4b | SWP07001 | 1.62 | -11.21 |
| BAT99 27 | HD 269333 | 05 18 19.27 | -69 11 40.6 | WN5b+B1 | SWP04830 | 1.36 | -10.69 |
| BAT99 29 | Brey 23 | 05 20 44.72 | -65 28 20.3 | WN4b | SWP09170 | 1.84 | -11.01 |
| BAT99 30 | Brey 24 | 05 21 57.59 | -65 48 59.3 | WN6h | SWP38382 | 1.43 | -10.94 |
| BAT99 31 | Brey 25 | 05 22 04.34 | -67 59 06.2 | WN3b | SWP33455 | 1.90 | -11.47 |
| BAT99 32 | HD 36063 | 05 22 22.54 | -71 35 57.8 | WN6(h) | SWP38235 | 1.33 | -10.72 |
| BAT99 33 | HD 269445 | 05 22 59.73 | -68 01 46.3 | WN9/Ofpe | SWP03001 | 0.80 | -11.69 |
| BAT99 35 | Brey 27 | 05 23 17.88 | -65 56 56.4 | WN3o | SWP23387 | 1.73 | -11.16 |
| BAT99 36 | HD 269485 | 05 24 23.97 | -68 31 35.3 | WN4b/WCE | SWP09160 | 2.01 | -10.99 |
| BAT99 41 | HD 269549 | 05 26 42.60 | -69 06 56.3 | WN4b | SWP38451 | 1.98 | -11.00 |
| BAT99 42 | HD 269546 | 05 26 45.29 | -68 49 51.4 | WN5b+(B3) | SWP08042 | 1.27 | -10.34 |
| BAT99 43 | Brey 37 | 05 27 37.71 | -70 36 05.4 | WN4+OB | SWP09161 | 1.46 | -11.19 |
| BAT99 44 | Brey 36 | 05 27 42.63 | -69 09 59.1 | WN8h | SWP06968 | 1.12 | -11.32 |
| BAT99 46 | Brey 38 | 05 28 17.93 | -69 02 34.7 | WN4o | SWP23385 | 1.73 | -11.73 |
| BAT99 48 | HD 269624 | 05 29 31.66 | -68 54 28.1 | WN4b | SWP39467 | 1.94 | -11.13 |
| BAT99 49 | Brey 40a | 05 29 33.16 | -70 59 35.2 | WN3+O6 | SWP38701 | 0.81 | -11.69 |
| BAT99 55 | HD 269687 | 05 31 25.49 | -69 05 37.9 | WN11h | SWP48710 | 0.68 | -11.52 |
| BAT99 56 | HD 269692 | 05 31 32.90 | -67 40 46.2 | WN4b | SWP38249 | 1.93 | -10.99 |
| BAT99 57 | Brey 45 | 05 31 34.51 | -67 16 28.7 | WN4b | SWP48032 | 1.14 | -11.78 |
| BAT99 58 | Brey 47 | 05 32 07.58 | -68 26 30.8 | WN6h | SWP48039 | 1.38 | -12.77 |
| BAT99 59 | HD 269748 | 05 33 10.61 | -67 42 42.7 | WN4o?+B | SWP09172 | 1.27 | -11.16 |
| BAT99 63 | Brey 52 | 05 34 52.10 | -67 21 28.4 | WN4h+abs | SWP09158 | 1.35 | -11.44 |
| BAT99 68 | Brey 58 | 05 35 42.21 | -69 11 54.2 | WN5-6 | SWP23891 | 1.37 | -12.20 |
| BAT99 75 | Brey 59 | 05 35 54.13 | -67 02 48.1 | WN3 | SWP09891 | 1.52 | -11.17 |
| BAT99 76 | Brey 64 | 05 35 54.43 | -68 59 07.1 | WN9h | SWP23890 | 0.64 | -12.36 |
| BAT99 79 | Brey 57 | 05 35 59.87 | -69 11 21.4 | WN7h+OB | SWP04895 | 0.68 | -11.52 |
| BAT99 81 | Brey 65a | 05 36 12.14 | -67 34 57.3 | WN5 | SWP40900 | 1.10 | -12.70 |
| BAT99 82 | Brey 66 | 05 36 33.61 | -69 09 16.8 | WN3b | SWP33622 | 1.16 | -12.11 |
| BAT99 89 | HD 269883 | 05 37 40.54 | -69 07 57.2 | WN7h | SWP48106 | 1.45 | -11.78 |
| BAT99 92 | HD 269891 | 05 37 49.01 | -69 05 07.8 | WN6+B1 | SWP04847 | 1.33 | -11.29 |
| BAT99 95 | Brey 80 | 05 38 33.69 | -69 04 49.9 | WN7h | SWP10701 | 1.46 | -11.07 |
| BAT99 118 | HD 38282 | 05 38 53.43 | -69 02 00.3 | WN6h | SWP04367 | 1.47 | -10.28 |
| BAT99 119 | HD 269928 | 05 38 57.14 | -69 06 05.2 | WN6(h) | SWP47825 | 1.23 | -11.29 |
| BAT99 120 | HD 269927c | 05 38 58.04 | -69 29 18.6 | WN9h | SWP48108 | 0.28 | -12.10 |
| BAT99 122 | HD 38344 | 05 39 11.38 | -69 02 01.0 | WN5(h) | SWP04139 | 1.52 | -11.22 |
| BAT99 132 | Brey 99 | 05 45 24.09 | -67 05 56.1 | WN4b | SWP23386 | 2.08 | -11.32 |
| BAT99 133 | SK -67 266 | 05 45 51.85 | -67 14 25.4 | WN11h | SWP48711 | 0.94 | -11.58 |
| BAT99 134 | HD 270149 | 05 46 46.27 | -67 09 57.5 | WN4b | SWP09912 | 2.07 | -11.03 |

a. W-R identifier from Breysacher et al. (1999)
b. alternate designation
c. Right Ascension (2000) from Breysacher et al. (1999)
d. Declination (2000) from Breysacher et al. (1999)
e. spectral type from Breysacher et al. (1999)
f. identifier of the IUE spectrum
g. measured equivalent width of He II 1640

h. measured line flux of He II 1640

Table 3. Optical data for all program stars.

| W-R ID[a] | log(*EW*(4686))[b] (Å) | log(*F*(4686))[c] (erg s$^{-1}$ cm$^{-2}$) | Reference[d] | *v*[e] (mag) | *(b−v)*[f] (mag) | Reference[g] | log($F_{calc}$(4686))[h] (erg s$^{-1}$ cm$^{-2}$) |
|---|---|---|---|---|---|---|---|
| WR 1 | 2.69 | -10.03 | 1 | 10.51 | 0.51 | 12 | -9.95 |
| WR 2 | 2.30 | -10.62 | 1 | 11.33 | 0.13 | 12 | -10.59 |
| WR 3 | 1.74 | -10.84 | 1 | 10.70 | -0.06 | 12 | -10.85 |
| WR 6 | 2.49 | -8.68 | 2 | 6.94 | -0.07 | 12 | -8.59 |
| WR 7 | 2.63 | -10.52 | 2 | 11.68 | 0.38 | 12 | -10.45 |
| WR 8 | 2.04 | -10.53 | 1 | 10.48 | 0.47 | 12 | -10.58 |
| WR 10 | 1.71 | -11.09 | 1, 2 | 11.08 | 0.22 | 12 | -11.10 |
| WR 12 | 1.87 | -10.96 | 1, 2 | 10.99 | 0.42 | 12 | -10.94 |
| WR 16 | 1.65 | -10.53 | 2 | 8.44 | 0.30 | 12 | -10.12 |
| WR 18 | 2.62 | | 3 | 11.11 | 0.55 | 12 | -10.26 |
| WR 21 | 1.71 | -10.57 | 1 | 9.76 | 0.27 | 12 | -10.58 |
| WR 22 | 1.42 | -9.61 | 1 | 6.44 | 0.03 | 12 | -9.49 |
| WR 24 | 1.49 | -9.60 | 1 | 6.49 | -0.06 | 12 | -9.42 |
| WR 25 | 1.22 | -10.59 | 1 | 8.14 | 0.17 | 12 | -10.40 |
| WR 31 | 1.81 | -10.88 | 2 | 10.69 | 0.28 | 12 | -10.85 |
| WR 40 | 1.68 | -9.67 | 1 | 7.85 | 0.11 | 12 | -9.81 |
| WR 43a | 1.20 | -12.13 | 2 | 11.90 | 0.74 | 12 | -12.03 |
| WR 44 | 2.35 | -11.27 | 2 | 12.96 | 0.37 | 12 | -11.24 |
| WR 46 | 2.02 | -10.65 | 1, 2 | 10.87 | -0.03 | 12 | -10.65 |
| WR 47 | 1.91 | -10.99 | 1 | 11.08 | 0.76 | 12 | -11.00 |
| WR 55 | 2.12 | -10.62 | 1, 2 | 10.87 | 0.40 | 12 | -10.64 |
| WR 58 | 2.51 | | 3 | 13.05 | 0.57 | 12 | -11.15 |
| WR 61 | 2.25 | -11.17 | 1, 2 | 12.41 | 0.36 | 12 | -11.12 |
| WR 71 | 2.20 | -10.26 | 1, 2 | 10.23 | -0.05 | 12 | -10.21 |
| WR 75 | 2.51 | -10.39 | 1 | 11.23 | 0.71 | 12 | -10.45 |
| WR 78 | 1.63 | | 4 | 6.61 | 0.21 | 12 | -9.39 |
| WR 85 | 1.95 | | 3 | 10.60 | 0.57 | 12 | -10.73 |
| WR 108 | 1.04 | -11.54 | 2 | 10.16 | 0.80 | 12 | -11.51 |
| WR 110 | 2.64 | -9.98 | 2 | 10.30 | 0.75 | 12 | -9.96 |
| WR 123 | 1.47 | | 3, 4 | 11.26 | 0.43 | 12 | -11.45 |
| WR 127 | 1.82 | -10.68 | 1 | 10.33 | 0.15 | 12 | -10.67 |
| WR 128 | 2.04 | -10.50 | 1 | 10.54 | -0.01 | 12 | -10.50 |
| WR 133 | 1.33 | -9.66 | 1 | 6.70 | 0.00 | 12 | -9.68 |
| WR 134 | 2.64 | -9.04 | 1 | 8.23 | 0.20 | 12 | -9.02 |
| WR 136 | 2.58 | -8.89 | 1 | 7.65 | 0.23 | 12 | -8.86 |
| WR 138 | 1.83 | -9.77 | 1 | 8.10 | 0.22 | 12 | -9.78 |
| WR 141 | 2.36 | -10.31 | 1 | 10.14 | 0.71 | 12 | -10.16 |
| WR 148 | 1.35 | -11.18 | 1 | 10.46 | 0.36 | 12 | -11.24 |
| WR 151 | 2.00 | -11.38 | 1 | 12.37 | 0.65 | 12 | -11.41 |
| WR 152 | 2.04 | -10.98 | 1 | 11.67 | 0.17 | 12 | -10.99 |
| WR 153 | 1.68 | -10.32 | 1 | 9.08 | 0.27 | 12 | -10.34 |
| WR 155 | 1.53 | -10.32 | 1 | 8.75 | 0.28 | 12 | -10.36 |
| WR 156 | 1.39 | | 3, 4 | 11.09 | 0.83 | 12 | -11.53 |
| WR 157 | 1.57 | | 3 | 9.91 | 0.46 | 12 | -10.82 |
| WR 158 | 1.58 | | 3 | 11.46 | 0.75 | 12 | -11.48 |
| BAT99 1 | 2.48 | -12.18 | 1, 5 | 15.70 | 0.01 | 13 | -12.13 |
| BAT99 2 | 2.20 | -12.59 | 6 | 16.22 | -0.16 | 14 | -12.57 |
| BAT99 3 | 2.41 | -11.76 | 1, 5 | 15.07 | -0.25 | 13 | -11.88 |
| BAT99 7 | 2.68 | -11.11 | 5 | 14.10 | -0.15 | 13 | -11.25 |
| BAT99 12 | 1.09 | | 7 | 13.72 | -0.23 | 14 | -12.67 |
| BAT99 13 | 0.41 | | 7, 11 | 12.98 | -0.02 | 14 | -13.10 |
| BAT99 14 | 1.94 | | 9 | 13.70 | 0.93 | 14 | -12.04 |
| BAT99 15 | 2.51 | -11.55 | 1, 5 | 14.83 | -0.07 | 13 | -11.73 |
| BAT99 16 | 1.93 | -11.51 | 1, 2 | 12.73 | -0.10 | 13 | -11.46 |
| BAT99 17 | 2.18 | -11.80 | 1, 2, 5 | 14.39 | -0.09 | 13 | -11.88 |
| BAT99 18 | 2.09 | -12.00 | 5 | 14.82 | -0.16 | 13 | -12.12 |
| BAT99 19 | 2.28 | | 3 | 13.76 | -0.06 | 13 | -11.53 |
| BAT99 22 | 0.84 | | 3, 7 | 12.09 | 0.07 | 13 | -12.34 |
| BAT99 24 | 2.57 | -11.51 | 5 | 14.74 | -0.08 | 13 | -11.63 |
| BAT99 26 | 2.37 | -11.77 | 5 | 14.71 | -0.11 | 13 | -11.81 |
| BAT99 27 | 1.20 | -11.56 | 1 | 11.31 | -0.14 | 13 | -11.62 |
| BAT99 29 | 2.37 | -11.80 | 1 | 14.62 | -0.08 | 13 | -11.78 |
| BAT99 30 | 1.88 | -11.76 | 2 | 13.40 | -0.13 | 13 | -11.77 |
| BAT99 31 | 2.49 | -12.10 | 1, 5 | 15.58 | 0.06 | 13 | -12.08 |
| BAT99 32 | 2.06 | -11.38 | 2 | 12.72 | -0.17 | 13 | -11.31 |
| BAT99 33 | 0.45 | | 7 | 11.54 | 0.11 | 14 | -12.52 |
| BAT99 35 | 2.24 | -11.96 | 5 | 14.72 | -0.01 | 13 | -11.97 |
| BAT99 36 | 2.48 | -11.72 | 1 | 14.80 | -0.08 | 13 | -11.75 |
| BAT99 41 | 2.39 | -11.82 | 5 | 15.00 | -0.11 | 13 | -11.91 |
| BAT99 42 | 1.11 | -11.20 | 1 | 9.91 | -0.09 | 13 | -11.16 |
| BAT99 43 | 2.01 | -11.85 | 1 | 14.18 | -0.21 | 13 | -11.94 |
| BAT99 44 | 1.59 | -12.04 | 1 | 13.47 | -0.19 | 13 | -12.08 |
| BAT99 46 | 2.17 | -12.27 | 1, 5 | 15.67 | 0.05 | 13 | -12.43 |
| BAT99 48 | 2.38 | -11.74 | 5 | 14.92 | -0.03 | 13 | -11.91 |
| BAT99 49 | 0.89 | -12.68 | 10 | 13.63 | -0.23 | 13 | -12.83 |
| BAT99 55 | -0.30 | | 7 | 11.99 | -0.16 | 14 | -13.38 |
| BAT99 56 | 2.29 | -11.80 | 5 | 14.74 | -0.07 | 13 | -11.91 |
| BAT99 57 | 2.42 | -11.80 | 5 | 15.19 | 0.09 | 13 | -12.00 |
| BAT99 58 | 1.64 | | 3, 7 | 15.10 | 0.23 | 13 | -12.78 |
| BAT99 59 | 1.63 | -11.88 | 1 | 13.33 | -0.21 | 13 | -11.98 |
| BAT99 63 | 1.86 | -12.24 | 1, 5 | 14.88 | -0.25 | 13 | -12.36 |
| BAT99 68 | 1.23 | | 3, 7 | 14.43 | 0.04 | 13 | -12.88 |
| BAT99 75 | 2.27 | -11.77 | 5 | 14.66 | -0.11 | 13 | -11.89 |
| BAT99 76 | 1.26 | -12.45 | 3, 7, 8 | 13.46 | 0.01 | 13 | -12.45 |
| BAT99 79 | 1.50 | | 3, 7 | 13.58 | 0.35 | 13 | -12.33 |
| BAT99 81 | 1.57 | -12.85 | 5 | 15.40 | 0.00 | 13 | -12.92 |
| BAT99 82 | 2.43 | -12.30 | 1, 5 | 16.12 | -0.03 | 13 | -12.34 |
| BAT99 89 | 1.82 | | 7 | 14.28 | -0.01 | 14 | -12.21 |
| BAT99 92 | 1.24 | | 1 | 11.51 | 0.03 | 13 | -11.70 |
| BAT99 95 | 2.19 | -11.36 | 1 | 13.16 | 0.13 | 14 | -11.43 |
| BAT99 118 | 1.95 | -10.90 | 1, 2 | 11.15 | -0.11 | 13 | -10.81 |
| BAT99 119 | 1.87 | -11.31 | 1, 2 | 12.16 | 0.03 | 13 | -11.33 |
| BAT99 120 | 0.71 | | 7, 8 | 12.59 | -0.18 | 14 | -12.61 |
| BAT99 122 | 1.92 | -11.54 | 1, 5 | 12.59 | 0.07 | 13 | -11.46 |
| BAT99 132 | 2.50 | -11.72 | 1, 5 | 15.02 | 0.07 | 13 | -11.85 |
| BAT99 133 | 0.04 | | 7 | 12.10 | -0.16 | 14 | -13.09 |
| BAT99 134 | 2.48 | -11.61 | 1, 5 | 14.63 | -0.04 | 13 | -11.69 |

a. W-R identifier from van der Hucht (2001) or Breysacher et al. (1999)
b. published equivalent width of He II 4686
c. published line flux of He II 4686
d. references for the 4686 data: (1) Conti & Morris (1990); (2) Smith et al. (1996); (3) Conti & Massey (1989); (4) Crowther et al. (1995b); (5) Crowther & Hadfield (2006); (6) P. Crowther, priv. comm.; (7) Schnurr et al. (2008); (8) Crowther et al. (1995a); (9) Foellmi et al. (2003); (10) Torres-Dodgen & Massey (1988); (11) Crowther & Smith (1997)
e. published *v* magnitude
f. published *(b–v)* color
g. references for *(b–v)*: (12) van der Hucht (2001); (13) Breysacher et al. (1999); (14) calculated from *(B–V)*
h. calculated line flux of He II 4686

Table 4. Reddening-free He II line fluxes

| W-R ID[a] | $E(b-v)$[b] (mag) | Reference[c] | $\log(F_0(1640))$[d] (erg s$^{-1}$ cm$^{-2}$) | $\log(F_0(4686))$[e] (erg s$^{-1}$ cm$^{-2}$) | $\log R_0$[f] |
|---|---|---|---|---|---|
| WR 1 | 0.67 | 1 | -7.70 | -8.78 | 1.08 |
| WR 2 | 0.44 | 1 | -8.87 | -9.80 | 0.93 |
| WR 3 | 0.35 | 1 | -9.15 | -10.19 | 1.04 |
| WR 6 | 0.12 | 1 | -7.30 | -8.45 | 1.15 |
| WR 7 | 0.53 | 1 | -8.55 | -9.54 | 0.98 |
| WR 8 | 0.53 | 2 | -8.95 | -9.55 | 0.59 |
| WR 10 | 0.58 | 1 | -8.93 | -10.01 | 1.08 |
| WR 12 | 0.80 | 1 | -8.39 | -9.47 | 1.08 |
| WR 16 | 0.55 | 1 | -8.07 | -9.50 | 1.43 |
| WR 18 | 0.75 | 1 | -7.75 | -8.86 | 1.11 |
| WR 21 | 0.51 | 2 | -8.83 | -9.62 | 0.78 |
| WR 22 | 0.35 | 1 | -7.66 | -8.96 | 1.30 |
| WR 24 | 0.24 | 1 | -8.06 | -9.15 | 1.09 |
| WR 25 | 0.63 | 1 | -7.69 | -9.42 | 1.73 |
| WR 31 | 0.55 | 2 | -8.75 | -9.85 | 1.10 |
| WR 40 | 0.40 | 1 | -8.16 | -8.93 | 0.78 |
| WR 43a | 1.02 | 3 | -8.08 | -10.24 | 2.16 |
| WR 44 | 0.62 | 1 | -8.81 | -10.12 | 1.31 |
| WR 46 | 0.30 | 1 | -9.18 | -10.10 | 0.91 |
| WR 47 | 0.83 | 2 | -8.77 | -9.45 | 0.69 |
| WR 55 | 0.65 | 1 | -8.48 | -9.40 | 0.92 |
| WR 58 | 0.43 | 2 | -9.71 | -10.35 | 0.64 |
| WR 61 | 0.55 | 1 | -9.31 | -10.14 | 0.83 |
| WR 71 | 0.38 | 1 | -8.74 | -9.55 | 0.82 |
| WR 75 | 0.93 | 1 | -8.14 | -8.65 | 0.51 |
| WR 78 | 0.47 | 1 | -7.46 | -8.51 | 1.06 |
| WR 85 | 0.82 | 1 | -8.25 | -9.21 | 0.96 |
| WR 110 | 0.90 | 1 | -7.35 | -8.30 | 0.95 |
| WR 123 | 0.75 | 1 | -9.21 | -10.05 | 0.84 |
| WR 127 | 0.46 | 2 | -8.96 | -9.82 | 0.86 |
| WR 128 | 0.32 | 1 | -8.90 | -9.90 | 1.00 |
| WR 133 | 0.40 | 2 | -7.92 | -8.92 | 1.00 |
| WR 134 | 0.47 | 1 | -7.37 | -8.16 | 0.79 |
| WR 136 | 0.45 | 1 | -7.41 | -8.06 | 0.65 |
| WR 138 | 0.57 | 2 | -7.76 | -8.71 | 0.95 |
| WR 141 | 1.05 | 2 | -7.63 | -8.36 | 0.73 |
| WR 148 | 0.83 | 1 | -8.65 | -9.64 | 0.99 |
| WR 151 | 0.82 | 2 | -9.19 | -9.86 | 0.67 |
| WR 152 | 0.50 | 1 | -8.92 | -10.04 | 1.12 |
| WR 153 | 0.56 | 2 | -8.56 | -9.27 | 0.71 |
| WR 155 | 0.53 | 2 | -8.79 | -9.34 | 0.55 |
| WR 156 | 1.22 | 1 | -8.12 | -9.25 | 1.13 |
| WR 157 | 0.88 | 2 | -8.47 | -9.19 | 0.72 |
| WR 158 | 1.08 | 1 | -8.34 | -9.46 | 1.12 |
| BAT99 1 | 0.14 | 4 | -11.15 | -12.18 | 1.03 |
| BAT99 2 | 0.13 | 4 | -11.83 | -12.59 | 0.76 |
| BAT99 3 | 0.12 | 4 | -10.75 | -11.76 | 1.01 |
| BAT99 7 | 0.08 | 4 | -10.13 | -11.11 | 0.98 |
| BAT99 12 | 0.10 | 4 | -11.59 | -12.67 | 1.08 |
| BAT99 14 | 0.09 | 4 | -10.80 | -12.04 | 1.24 |
| BAT99 15 | 0.08 | 4 | -10.53 | -11.55 | 1.02 |
| BAT99 16 | 0.09 | 4 | -10.53 | -11.51 | 0.98 |
| BAT99 17 | 0.11 | 4 | -10.76 | -11.80 | 1.04 |
| BAT99 18 | 0.10 | 4 | -10.98 | -12.00 | 1.02 |
| BAT99 19 | 0.16 | 4 | -10.52 | -11.53 | 1.01 |
| BAT99 24 | 0.10 | 4 | -10.55 | -11.51 | 0.96 |
| BAT99 26 | 0.14 | 4 | -10.97 | -11.77 | 0.80 |
| BAT99 27 | 0.23 | 4 | -10.46 | -11.56 | 1.10 |
| BAT99 29 | 0.12 | 4 | -10.78 | -11.80 | 1.02 |
| BAT99 30 | 0.07 | 4 | -10.71 | -11.76 | 1.05 |
| BAT99 31 | 0.17 | 4 | -11.24 | -12.10 | 0.86 |
| BAT99 32 | 0.08 | 4 | -10.49 | -11.38 | 0.89 |
| BAT99 35 | 0.11 | 4 | -10.93 | -11.96 | 1.03 |
| BAT99 36 | 0.13 | 4 | -10.76 | -11.72 | 0.96 |
| BAT99 41 | 0.12 | 4 | -10.77 | -11.82 | 1.05 |
| BAT99 42 | 0.30 | 4 | -10.11 | -11.20 | 1.09 |
| BAT99 43 | 0.13 | 4 | -10.96 | -11.85 | 0.89 |
| BAT99 44 | 0.12 | 4 | -11.09 | -12.04 | 0.95 |
| BAT99 46 | 0.21 | 4 | -11.50 | -12.27 | 0.77 |
| BAT99 48 | 0.10 | 4 | -10.90 | -11.74 | 0.84 |
| BAT99 49 | 0.15 | 4 | -11.46 | -12.68 | 1.22 |
| BAT99 56 | 0.12 | 4 | -10.76 | -11.80 | 1.04 |
| BAT99 57 | 0.10 | 4 | -11.55 | -11.80 | 0.25 |
| BAT99 58 | 0.50 | 4 | -12.54 | -12.78 | 0.24 |
| BAT99 59 | 0.16 | 4 | -10.93 | -11.88 | 0.95 |
| BAT99 63 | 0.10 | 4 | -11.21 | -12.24 | 1.03 |
| BAT99 68 | 0.52 | 4 | -11.96 | -12.88 | 0.92 |
| BAT99 75 | 0.07 | 4 | -10.94 | -11.77 | 0.83 |
| BAT99 79 | 0.50 | 4 | -11.29 | -12.33 | 1.04 |
| BAT99 81 | 0.33 | 4 | -12.47 | -12.85 | 0.38 |
| BAT99 82 | 0.27 | 4 | -11.88 | -12.30 | 0.42 |
| BAT99 89 | 0.28 | 4 | -11.55 | -12.21 | 0.66 |
| BAT99 92 | 0.39 | 4 | -11.06 | -11.70 | 0.64 |
| BAT99 95 | 0.25 | 4 | -10.84 | -11.36 | 0.52 |
| BAT99 118 | 0.16 | 4 | -10.05 | -10.90 | 0.85 |
| BAT99 119 | 0.29 | 4 | -11.06 | -11.31 | 0.25 |
| BAT99 122 | 0.28 | 4 | -10.99 | -11.54 | 0.55 |
| BAT99 132 | 0.23 | 4 | -11.09 | -11.72 | 0.63 |
| BAT99 134 | 0.06 | 4 | -10.80 | -11.61 | 0.81 |

a. W-R identifier from van der Hucht (2001) or Breysacher et al. (1999)
b. published reddening from continuum fitting
c. references for the reddening data: (1) Hamann et al. (2006); (2) Schmutz & Vacca (1991); (3) Crowther & Dessart (1998); (4) Hainich et al. (2014)
d. dereddened He II 1640 line flux
e. dereddened He II 4686 line flux
f. logarithmic ratio of the 1640/4686 flux

Table 5. Parameters for several interstellar reddening laws.

| Reddening Law[a] | $R_V$[b] | $k(1640)$[c] | $k(4686)$[d] | A[e] | B[f] | Reference[g] |
|---|---|---|---|---|---|---|
| Milky Way | 3.1 | 8.036 | 3.843 | -0.597 | 0.531 | Mathis (1990) |
| LMC | 3.41 | 8.886 | 4.149 | -0.528 | 0.470 | Gordon et al. (2003) |
| SMC | 2.74 | 11.548 | 3.498 | -0.310 | 0.276 | Gordon et al. (2003) |
| Local Galaxies | 4.05 | 9.839 | 4.767 | -0.493 | 0.439 | Calzetti et al. (2000) |
| MOSDEF Survey | 2.51 | 8.248 | 3.215 | -0.497 | 0.442 | Reddy et al. (2015) |

a. type of reddening law
b. adopted ratio of total to selective absorption in the *V* band
c. absorption coefficient at 1640 Å
d. absorption coefficient at 4686 Å
e. first order coefficient in equation (7)
f. zeroth order coefficient in equation (7)
g. reference for the reddening law

Table 6. Dust measurements for NGC 3049 and Tol 89 using different methods.

| | NGC 3049 ("B") | Tol 89 ("A1") | Tol 89 ("A2") | Tol 89 ("A1+A2") |
|---|---|---|---|---|
| $F(1640)$ ($10^{-15}$ erg s$^{-1}$ cm$^{-2}$) | 9.80 (+1.88, -1.73) | 4.89 (+1.39, -1.34) | 3.98 (+0.71, -0.80) | 8.87 (+1.52, -1.56) |
| $F(\mathrm{Bump})$ ($10^{-15}$ erg s$^{-1}$ cm$^{-2}$) | 10.83 (+0.49, -0.52) | 4.24 (+0.31, -0.35) | 2.32 (+0.24, -0.23) | 5.88 (+0.40, -0.40) |
| $F(4686)$ ($10^{-15}$ erg s$^{-1}$ cm$^{-2}$) | 3.38 (+0.24, -0.26) | 1.42 (+0.17, -0.19) | 0.65 (+0.11, -0.11) | 2.07 (+0.20, -0.22) |
| $F(1640)/F(4686)$ | 2.94 (+0.54, -0.53) | 3.53 (+1.08, -1.07) | 5.96 (+1.66, -1.35) | 4.23 (+0.87, -0.80) |
| $F(\mathrm{H}\alpha)$ ($10^{-15}$ erg s$^{-1}$ cm$^{-2}$) | 26.36 (+0.47, -0.51) | 3.30 (+0.64, -0.63) | 0.79 (+0.18, -0.17) | 4.20 (+0.69, -0.68) |
| $F(\mathrm{H}\beta)$ ($10^{-15}$ erg s$^{-1}$ cm$^{-2}$) | 6.58 (+0.26, -0.28) | 0.73 (+0.20, -0.24) | 0.25 (+0.15, -0.12) | 1.00 (+0.22, -0.21) |
| $F(\mathrm{H}\alpha)/F(\mathrm{H}\beta)$ | 4.00 (+0.17, -0.18) | 4.72 (+1.22, -1.05) | 3.01 (+2.54, -1.30) | 4.16 (+1.34, -0.94) |
| $\beta$ | -1.36 (+0.05, -0.05) | -1.73 (+0.07, -0.08) | -2.23 (+0.16, -0.17) | -1.81 (+0.07, -0.07) |
| $E(B-V)_{\mathrm{He\,II}}$ (Reddy) | 0.21 (+0.04, -0.04) | 0.17 (+0.08, -0.06) | 0.06 (+0.06, -0.05) | 0.13 (+0.04, -0.04) |
| $E(B-V)_{\mathrm{He\,II}}$ (LMC) | 0.22 (+0.05, -0.04) | 0.18 (+0.09, -0.06) | 0.06 (+0.06, -0.06) | 0.14 (+0.05, -0.04) |
| $E(B-V)_{\mathrm{Balmer}}$ (Reddy) | 0.32 (+0.04, -0.04) | 0.46 (+0.20, -0.21) | 0.05 (+0.52, -0.46) | 0.35 (+0.24, -0.22) |
| $E(B-V)_{\mathrm{Balmer}}$ (LMC) | 0.26 (+0.03, -0.04) | 0.39 (+0.34, -0.25) | 0.05 (+0.47, -0.42) | 0.29 (+0.22, -0.20) |
| $E(B-V)_{\mathrm{UV\,slope}}$ (Reddy) | 0.24 (+0.01, -0.01) | 0.16 (+0.02, -0.02) | 0.05 (+0.03, -0.04) | 0.14 (+0.02, -0.01) |